\documentclass[12pt,preprintnumbers]{article}
\usepackage{titling}
\usepackage{amsmath}
\usepackage{slashed}
\usepackage{amssymb}
\usepackage{epsfig}
\usepackage{graphicx}
\usepackage{multirow,booktabs}
\usepackage{tabu}
\usepackage{color}
\usepackage{rotating}
\usepackage{hyperref,afterpage}
\usepackage[margin=1.0in]{geometry}
\usepackage[usenames,dvipsnames,svgnames,table]{xcolor}
\usepackage{enumitem}
\usepackage[utf8x]{inputenc}
\usepackage[compress,numbers,sort]{natbib}
\usepackage{authblk}
\usepackage{colortbl}
\usepackage{pdflscape}
\usepackage{comment}
\usepackage{braket}
\usepackage{color}
\usepackage{subcaption}
 \usepackage{pgfplotstable}
\usepackage{array}
\usepackage{colortbl}
\usepackage{float}
\usepackage{subfloat}
\usepackage{siunitx}
\usepackage{placeins}
\definecolor{Gray}{gray}{0.95}

\usepackage{dsfont}
\usepackage{slashed}
\usepackage[normalem]{ulem}

\usepackage{titling}

\usepackage[compat=1.1.0]{tikz-feynman}
\pgfmathsetmacro\sizedot{1.1}
\pgfmathsetmacro\sizesqdot{1.6}
\pgfmathsetmacro\sizeemdot{2.}

\pgfmathsetmacro\lwR{.4}
\pgfmathsetmacro\lwL{1.2}

\usepackage{xparse}
\ExplSyntaxOn
\NewDocumentCommand{\longdash}{ O{2} }
{
	--\prg_replicate:nn { #1 - 1 } { \negthinspace -- }
}
\ExplSyntaxOff

\definecolor{nicered}{rgb}{0.6,0.1,0.1}
\definecolor{nicegreen}{rgb}{0.1,0.5,0.1}
\definecolor{mediumcandyapplered}{rgb}{0.99, 0.12, 0.07}
\definecolor{red}{rgb}{1.0, 0, 0}
\hypersetup{colorlinks,citecolor= nicegreen,linkcolor= nicered}

\newcommand{\coeff}[2]{ \mathcal{C}_{#1} ^{#2} }
\newcommand{\op}[2]{ \mathcal{O}_{#1} ^{#2} }
\newcommand{\ct}[2]{ \delta_{#1} ^{#2} }
\newcommand{\id}[1]{ \mathbf{1}_{#1} }

\newcommand{\order}[1] {\mathcal{O }\left( #1 \right)}
\newcommand{\yuk}[1]{{Y}_{#1}}
\newcommand{\hc}{\text{H.c.}}

\newcommand{\preprint}{%
  \begin{flushright}%
   \small{COMETA-2024-19}
  \end{flushright}%
  \vspace{2em}%
}

\title{
\preprint
\bf{Two-loop running effects in Higgs physics in Standard Model Effective Field Theory}
}
\author[1]
{Stefano Di Noi \thanks{stefano.dinoi@studenti.unipd.it}}
\affil[1]{\emph{\normalsize Dipartimento di Fisica e Astronomia ``G. Galilei", Universit\`a di Padova, and Istituto Nazionale di Fisica Nucleare, Sezione di Padova, I-35131 Padova, Italy}}
\author[1]
{Ramona Gr\"{o}ber\thanks{ramona.groeber@pd.infn.it}}
\author[1]
{Manoj K. Mandal \thanks{manojkumar.mandal@pd.infn.it}}

\date{}

\begin{document} 

\maketitle

\begin{abstract}
\normalsize
We consider the renormalization group equations within the Standard Model Effective Field Theory
and compute two-loop contributions proportional to the top quark Yukawa coupling
for the operator generating an effective Higgs-gluon coupling, focusing on the Yukawa-like operator.
These two-loop  running effects are relevant for processes where the
effective Higgs-gluon coupling contributes at a lower loop order compared to the Standard Model contribution and where a dynamical
scale choice is adopted. Such a situation arises, for instance, in the Higgs
transverse momentum distribution and Higgs pair production.
We investigate the phenomenological impact of our computations on
these two processes and find that the two-loop contributions are
significant and can lead to deviations of up to 20\% in the scenarios we consider.
\end{abstract}



\newpage
\section{Introduction}
There are various reasons that suggest that the Standard Model (SM) of particle physics should be extended. So far no striking direct evidence has been observed, motivating an effective field theory (EFT) approach to describe new physics as model-independently as possible. This approach is valid under the assumption that the new physics scale is much above the electroweak scale. Assuming that the Higgs boson transforms as in the SM in an $SU(2)_L$ doublet and writing down all possible higher-dimensional operators leads to the so-called SM Effective Field Theory (SMEFT). 
For collider physics, the most relevant effects arise at mass dimension six, for which a non-redundant basis has been given for the first time in Ref.~\cite{dim6smeft}. \par

At the LHC, a comprehensive programme of SMEFT analyses, both inclusive and differential, is conducted to investigate potential new physics or, at least, establish bounds on the new physics scale. While model-independent tests of new physics can also be performed using pseudo-observables \cite{Bardin:1999gt, Passarino:2010qk, Ghezzi:2015vva, Gonzalez-Alonso:2014eva, Greljo:2015sla, Gainer:2018qjm}, SMEFT enables a global analysis by correlating various experimental measurements across different energy scales.

An essential aspect of these analyses is hence to correctly relate the Wilson coefficients at different scales via renormalisation group equation (RGE) running. The complete one-loop running for dimension-6 operators has been computed in \cite{rge1, rge2, rge3}, and initial efforts towards two-loop SMEFT RGEs are reported in \cite{Bern:2019wie, Bern:2020ikv, Jenkins:2023bls}. Several one-loop level RGE implementations \cite{Lyonnet:2013dna, Celis:2017hod, Aebischer:2018bkb, Fuentes-Martin:2020zaz, DiNoi:2022ejg} are available. They are based on numerical solutions to solve the coupled system of the RGEs. 

These implementations are vital for global analyses involving observables at varying energy scales, such as combining low and high energy data.
Moreover, with the increasing precision of the LHC experiments the RGE effects turn out also to be sizeable when considering processes with dynamical renormalisation scale choice, where the relevant scale covers a large range.
In this context, the RGE running of the Wilson coefficients should be included, as was shown in \cite{Aoude:2022aro, Maltoni:2024dpn, Grazzini:2018eyk, Battaglia:2021nys}. Those works were focusing on the RGE running proportional to the strong coupling constant $\alpha_s$ with the advantage that the coupled system of RGEs can then be solved analytically. 
More recently, it has also been shown that the running proportional to the top quark Yukawa coupling $Y_t$ can have a sizeable effect \cite{DiNoi:2023onw}. 
Although the top quark Yukawa coupling contribution $\alpha_t=Y_t^2/4\pi$ is smaller than $\alpha_s$ the difference is overcome due to the fact that the Wilson coefficients that run with $\yuk{t}$ are generically less constrained \cite{Celada:2024mcf}, amplifying the RGE running effect. 
\par
In this paper, we focus on the impact of RGE running in Higgs physics. The dominant single and multi-Higgs production processes occur in the SM via a loop of heavy quarks, hence arises for the first time at one-loop level.
In the SMEFT, the situation is though different. The operator $\op{HG}{}=H^{\dagger} H G_{\mu\nu}G^{\mu\nu}$, where $H$ denotes the Higgs doublet and $G_{\mu\nu}$ the gluon field strength, generates an effective coupling of gluons to Higgs bosons, which leads to a tree-level contribution to Higgs production. 

The diagrams arising at the one-loop level, among which the SM contribution, get corrected by the RGE effect. When one-loop RGE running is employed, this counts effectively as a two-loop contribution. For consistency, this requires that the effective coupling of the Higgs boson to gluons arising at tree-level needs to include two-loop order RGEs.
\par
In weakly interacting models, a Wilson coefficient $\coeff{HG}{}$ to the effective Higgs gluon coupling is loop-generated \cite{Arzt:1994gp}. 
The operator mixing at one-loop level of $\op{HG}{}$ again stems from operators that (in the same assumption of weakly interacting UV models) are generated at one-loop level (as $\op{HG}{}$ itself or the chromomagnetic operator $\,\op{tG}{}$). In conclusion, as also shown in \cite{Grojean:2024tcw} even at dimension-8 level, $\op{HG}{}$  does not get renormalised at one-loop level.
In a consistent loop-counting this requires the consideration of the two-loop contributions to the RGEs of the tree-level-generated operators to remain consistent in the loop counting. In this paper, we take a step forward in this respect computing a missing piece proportional to the top quark Yukawa coupling in the two-loop RGE of $\coeff{HG}{}$. The terms that are proportional to the four-top quark operators have been computed in Ref.~\cite{DiNoi:2023ygk} and its interplay in two different continuation schemes for $\gamma_5$ with the chromomagnetic operator $\op{tG}{}$ is discussed. Indeed the continuation scheme dependence is unphysical and disappears when consistently matching to UV models \cite{DiNoi:2023ygk}.

The chromomagnetic operator, according to Ref.~\cite{Arzt:1994gp, Buchalla:2022vjp}, is loop-generated: four-top operators and the chromomagnetic operator arise at the same order in the running of $\coeff{HG}{}$. The scheme dependent contribution to $\op{tG}{}$ cancels the one of the four-top operators. 
This confirms that it is necessary to include the RGE effects of the potentially tree-level generated operators at two-loop level into the RGEs of the one-loop generated operators. 

We take a first step towards the complete two-loop running of $\coeff{HG}{}$ by computing the two-loop contribution stemming from the Yukawa-like operator $\op{t H}{} = (\bar{Q}_L \tilde{H} t_R ) \left( H^\dagger H \right)$ due to its phenomenological relevance in Higgs physics. We demonstrate that, although $\coeff{tH}{}$ is already constrained by Higgs measurements, including its two-loop RGE running still has a significant impact on the processes we consider.

Those are chosen such that the situation is as described before: 
\begin{enumerate}
    \item a wide range of energy scales is typically considered, 
    \item the $\coeff{HG}{}$ Wilson coefficients enters at tree-level while the SM contribution arises at the first time at one-loop level.
\end{enumerate} 
In this respect, we study the Higgs $p_{T,h}$ spectrum in Higgs+jet production which is of large phenomenological interest as it allows to resolve degeneracies from inclusive Higgs production \cite{Grojean:2013nya, Azatov:2013xha}. The second process we consider is Higgs pair production that provides a measurement of the trilinear Higgs self-coupling \cite{ATLAS:2022vkf, CMS:2022dwd} and can probe Higgs non-linearities \cite{Grober:2010yv, Contino:2012xk, Grober:2016wmf}. 
Both processes can bound light quark Yukawa couplings \cite{Bishara:2016jga, Soreq:2016rae, Alasfar:2019pmn, Alasfar:2022vqw}.

The paper is structured as follows. 
In Section~\ref{section:notation} we present our notation and convention for SMEFT operators and SM parameters. 
In Section~\ref{section:CHG} we present our computation of the two-loop contribution of the Yukawa-like operator to the running of the Higgs-gluon coupling. The phenomenological impact is studied for the Higgs $p_{T,h}$ spectrum and Higgs pair production in Section~\ref{section:pheno}. In Section \ref{section:conclusion} we present our conclusions. For completeness, we provide in Appendix \ref{sec:FeynmanRules} the Feynman rules we used and in Appendix \ref{sec:MI} the master integrals.
\section{Notation} \label{section:notation}
We work in the SMEFT, where all SM fields transform under the unbroken SM gauge group. 
The SMEFT Lagrangian can be written as an expansion in $1/\Lambda$
\begin{equation}
\mathcal{L}=\mathcal{L}_{\text{SM}}+\sum_{\mathcal{D}_i=5} \frac{\coeff{i}{}}{\Lambda} \op{i}{}
+\sum_{\mathcal{D}_i=6} \frac{\coeff{i}{}}{\Lambda^2} \op{i}{}+...
\end{equation}
where $\Lambda$ is the new physics scale and the sums run over all operators invariant under the SM gauge symmetries. With $\op{i}{}$ we denote generically the operator and with $\coeff{i}{}$ the associated Wilson coefficient. We follow the \textit{Warsaw basis} of Ref.~\cite{dim6smeft} that is a complete basis while removing all redundant operators.
Since we are interested in Higgs physics, we consider only operators with quarks of the third generation, hence slightly modify the original notation of Ref.~\cite{dim6smeft} specifying the third generation by $Q_L$, $t_R$ and $b_R$.
The operators we consider are
\begin{equation}
\begin{split}
\mathcal{L}_{\mathrm{\mathcal{D}=6}}&=
\frac{\coeff{H \Box}{}}{\Lambda^2} \op{H \Box}{}+ \frac{\coeff{H D}{}}{\Lambda^2} \op{H D}{}+
\frac{\coeff{H G}{}}{\Lambda^2} \op{H G}{}+   
\frac{\coeff{Qt(1)}{}}{\Lambda^2} \op{Qt(1)}{}+
\frac{\coeff{Qt(8)}{}}{\Lambda^2} \op{Qt(8)}{}+\frac{\coeff{H}{}}{\Lambda^2} \op{H}{} \\ &+
\left[\frac{\coeff{t H  }{}}{\Lambda^2} \op{t H}{}+\frac{\coeff{b H  }{}}{\Lambda^2} \op{b H}{} + \frac{\coeff{tG}{}}{\Lambda^2} \op{tG}{}+
\hc \right],
\end{split}
\end{equation}
with
\begin{equation}
\op{H \Box}{}=(H^{\dagger} H) \Box (H^{\dagger} H), \quad \op{HD}{}=(H^{\dagger} D_{\mu} H)^* (H^{\dagger} D^{\mu} H), \quad   \op{H}{}=(H^\dagger H)^3,
\end{equation}
\begin{equation}
\op{t H}{} = (\bar{Q}_L \tilde{H} t_R ) \left( H^\dagger H \right),\quad \op{b H}{} = (\bar{Q}_L {H} b_R ) \left( H^\dagger H \right), 
\end{equation}
\begin{equation}
\op{H G}{} = \left( H^\dagger H \right) G_{\mu \nu}^A G^{\mu \nu,A}, \quad \op{tG}{}=\bar{Q}_L \tilde{H} \sigma_{\mu\nu} T^A t_R G^{\mu\nu, A},
\end{equation}
\begin{equation}
\op{Qt(1)}{} = \left(\bar{Q}_L \gamma^\mu Q_L \right) \left(\bar{t}_R \gamma_\mu t_R \right), \quad
\op{Qt(8)}{} = \left(\bar{Q}_L \gamma^\mu T^A Q_L \right) \left(\bar{t}_R \gamma_\mu T^A  Q_R \right)\,. \label{eq:4top}
\end{equation}
In the previous expression we denote with $Q_L$ the $SU(2)_{L}$ doublet of the third quark generation, $t_R$ ($b_R$) for the right-handed top quark (bottom quark) field. The color generators are represented by $T^A$ while $\tau^I$ are the Pauli matrices. $G_{\mu\nu}^A=\partial_\mu G_\nu^A-\partial_\nu G_\mu^A-g_s f^{ABC}G_\mu^B G_\nu^C$ is the gluon field strength tensor, $H$ the usual scalar doublet field with $H=1/\sqrt{2}(0,v+h)$ in the unitary gauge and $\sigma_{\mu\nu}=i/2[\gamma_{\mu},\gamma_{\nu}]$. Moreover, $\tilde{H}_j= \epsilon_{jk} (H^k)^*$ with $\epsilon_{12}=+1$. We assume all the Wilson coefficients to be real, since we do not consider CP-violating effects.\footnote{For CP-violating operators in Higgs pair production, see \cite{Grober:2017gut}.}
We note that the list of four-fermion operators is not complete but we denote in Eq.~\eqref{eq:4top} only the ones that contribute to the RGE running of $\coeff{HG}{}$ \cite{DiNoi:2023ygk}.

The processes $pp \to hh,hj$ are affected by: 
\begin{enumerate}
    \item  The operator $\op{HG}{}$ that introduces a tree-level coupling of gluons to one and two Higgs bosons. 
    \item The operator $\op{tH}{}$ that changes the SM top Yukawa coupling and generates an effective coupling of two Higgs bosons to two top quarks.
    \item The operator $\op{tG}{}$ that modifies the gluon top quark coupling and generates an effective coupling of a Higgs boson to gluons and top quarks.
    \item The operators $\op{H\Box}{}$ and $\op{HD}{}$ that generate an overall rescaling of the SM couplings as they require a field-redefinition for a canonically normalised Higgs propagator. 
    \item The operator $\op{H}{}$ that changes the trilinear Higgs self-coupling affects at leading order (LO) Higgs pair production only.
\end{enumerate}
Some illustrative Feynman diagrams can be found in Fig.~\ref{fig:diagramshj} for $hj$ production and in Fig.~\ref{fig:diagramshh} for $hh$ production.
\begin{figure}[t!]
\centering
    \begin{subfigure}{0.3\textwidth}
        \centering
        \begin{tikzpicture} 
            \begin{feynman}[small]
                \vertex  (g1)  {$g$};
                \vertex (HG) [dot, scale=0.01,below right= of g1,color=white] {};
                \vertex (g2) [below left = of HG]  {$g$};
                \vertex (h) [above right = of HG] {$h$};
                \vertex (j) [below right = of HG] {$g$};
                \diagram* {
                    (g1)  -- [gluon] (HG),
                    (g2) -- [gluon] (HG),
                    (h)  -- [scalar] (HG),
                     (HG) -- [gluon] (j)
                };
             \node[shape=regular polygon,regular polygon sides=4, aspect=1,fill=white, draw=black,scale = .5,opacity=1.] at (HG) {};  
            \end{feynman}
        \end{tikzpicture}
    \end{subfigure}
\begin{subfigure}{0.3\textwidth}
\centering
        \begin{tikzpicture} 
            \begin{feynman}[small]
                \vertex  (g1)  {$g$};
                \vertex  (gtt1) [dot,scale=\sizedot,right = of g1] {};
                \vertex (CM) [square dot, scale=\sizesqdot,below right= of gtt1,color=black] {};
                \vertex  (gtt2) [dot,scale=\sizedot,below left = of CM] {};
                \vertex (g2) [left = of gtt2]  {$g$};
                \vertex (h) [above right = of CM] {$h$};
                \vertex (j) [below right = of CM] {$g$};
                \diagram* {
                    (g1)  -- [gluon] (gtt1),
                    (g2) -- [gluon] (gtt2),
                    (h)  -- [scalar] (CM),
                    (gtt1) -- [fermion] (CM) 
                    -- [fermion] (gtt2)
                     -- [fermion] (gtt1), (CM) -- [gluon] (j)
                };
                \node[shape=regular polygon,regular polygon sides=4, aspect=1,fill=black, draw=black,scale = .5,opacity=1.] at (CM) {};  
            \end{feynman}
        \end{tikzpicture}
    \end{subfigure}
    \begin{subfigure}{0.3\textwidth}
        \centering
       \begin{tikzpicture} 
            \begin{feynman}[small]
                \vertex  (g1)  {$g$};
                \vertex  (gtt1) [dot,scale=\sizedot,right =  of g1] {};
                \vertex (tH) [square dot, scale=\sizesqdot,right=35 pt of gtt1,color=black] {};
                \vertex  (gtt2) [dot,scale=\sizedot,below =35 pt of gtt1] {};
                 \vertex  (gtt3) [dot,scale=\sizedot,right = 35 pt of gtt2] {};
                \vertex (g2) [left = of gtt2]  {$g$};
                \vertex (h) [right = of tH] {$h$};
                \vertex (j) [right = of gtt3] {$g$};
                \diagram* {
                    (g1)  -- [gluon] (gtt1),
                    (g2) -- [gluon] (gtt2),
                    (h)  -- [scalar] (tH),
                    (gtt1) -- [fermion] (tH) -- [fermion] (gtt3) 
                    -- [fermion] (gtt2)
                     -- [fermion] (gtt1), 
                     (gtt3) -- [gluon] (j),
                };
                 \node[shape=regular polygon,regular polygon sides=3, aspect=1,fill=white, draw=black,scale = .5,opacity=1.] at (tH) {};  
            \end{feynman}
        \end{tikzpicture}
    \end{subfigure}
\caption{Sample of the diagrams contributing to $pp \to hj$ with an insertion of the operators $\op{HG}{}$ (white square), $\op{tG}{}$ (black square) and $\op{tH}{}$ (white triangle).} \label{fig:diagramshj}
\end{figure}
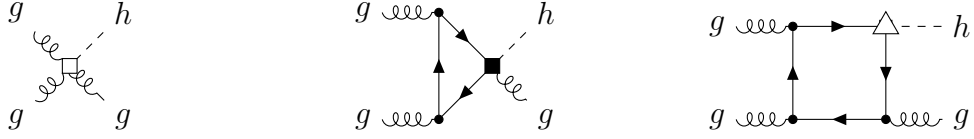
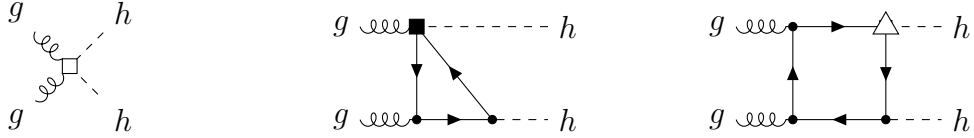
\begin{figure}[t!]
\centering
    \begin{subfigure}{0.3\textwidth}
        \centering
        \begin{tikzpicture} 
            \begin{feynman}[small]
                \vertex  (g1)  {$g$};
                \vertex (HG) [dot, scale=0.01,below right= of g1,color=white] {};
                \vertex (g2) [below left = of HG]  {$g$};
                \vertex (h) [above right = of HG] {$h$};
                \vertex (j) [below right = of HG] {$h$};
                \diagram* {
                    (g1)  -- [gluon] (HG),
                    (g2) -- [gluon] (HG),
                    (h)  -- [scalar] (HG),
                     (HG) -- [scalar] (j)
                };
             \node[shape=regular polygon,regular polygon sides=4, aspect=1,fill=white, draw=black,scale = .5,opacity=1.] at (HG) {};  
            \end{feynman}
        \end{tikzpicture}
    \end{subfigure}
\begin{subfigure}{0.3\textwidth}
\centering
\begin{tikzpicture} 
            \begin{feynman}[small]
                \vertex  (g1)  {$g$};
                \vertex  (gtt1) [square dot, scale=0.01,right= of g1,color=blue] {};
                \vertex  (gtt2)                 [dot,scale=\sizedot,below = 35 pt  of gtt1] {};
                \vertex  (g2) [left=of gtt2]  {$g$};
                \vertex (htt) [right = of gtt2, dot, scale=\sizedot] {};
                \vertex (h2) [right = of htt] {$h$};
                \vertex (h1) [above = 35 pt  of h2] {$h$};
                \diagram* {
                    (g1)  -- [gluon] (gtt1),
                    (g2) -- [gluon] (gtt2),
                    (h1)  -- [scalar] (gtt1),
                    (gtt1) -- [fermion] (gtt2) -- [fermion] (htt)
                     -- [fermion] (gtt1), 
                    (htt) -- [scalar] (h2),
                };
            \end{feynman}
             \node[shape=regular polygon,regular polygon sides=4, aspect=1,fill=black, draw=black,scale = .5,opacity=1.] at (gtt1) {};  
        \end{tikzpicture}
    \end{subfigure}
    \begin{subfigure}{0.3\textwidth}
        \centering
       \begin{tikzpicture} 
            \begin{feynman}[small]
                \vertex  (g1)  {$g$};
                \vertex  (gtt1) [dot,scale=\sizedot,right =  of g1] {};
                \vertex (tH) [square dot, scale=\sizesqdot,right=35 pt of gtt1,color=black] {};
                \vertex  (gtt2) [dot,scale=\sizedot,below =35 pt of gtt1] {};
                 \vertex  (gtt3) [dot,scale=\sizedot,right = 35 pt of gtt2] {};
                \vertex (g2) [left = of gtt2]  {$g$};
                \vertex (h) [right = of tH] {$h$};
                \vertex (j) [right = of gtt3] {$h$};
                \diagram* {
                    (g1)  -- [gluon] (gtt1),
                    (g2) -- [gluon] (gtt2),
                    (h)  -- [scalar] (tH),
                    (gtt1) -- [fermion] (tH) -- [fermion] (gtt3) 
                    -- [fermion] (gtt2)
                     -- [fermion] (gtt1), 
                     (gtt3) -- [scalar] (j),
                };
                 \node[shape=regular polygon,regular polygon sides=3, aspect=1,fill=white, draw=black,scale = .5,opacity=1.] at (tH) {};  
            \end{feynman}
        \end{tikzpicture}
    \end{subfigure}
\caption{Sample of the diagrams contributing to $pp \to hh$ with an insertion of the operators $\op{HG}{}$ (white square), $\op{tG}{}$ (black square) and $\op{tH}{}$ (white triangle).} \label{fig:diagramshh}
\end{figure}
For completeness, we give also our notation for the SM Lagrangian
\begin{equation}\label{eq:SMlag}
\begin{aligned}
\mathcal{L}_{\textrm{SM}} =&  -\frac{1}{4} G_{\mu \nu}^A G^{A\mu \nu}
			-\frac{1}{4} W_{\mu \nu}^I W^{I\mu \nu}
			-\frac{1}{4} B_{\mu \nu}B^{\mu \nu} + \sum_{\psi} \bar{\psi} i \slashed{D} \psi 
			+ (D_\mu H)^\dagger (D^\mu H) \\ & + \mu^2 H^\dagger H -\lambda (H^\dagger H)^2   - \left[
 \yuk{t} \tilde{H}^\dagger \bar{t}_R  Q_L+\yuk{b}  {H}^\dagger \bar{b}_R  Q_L + \hc\right].
\end{aligned}
\end{equation}
In the previous expression, $\psi = Q_L, t_R, b_R$. 
We use the plus sign covariant derivative, namely $D_{\mu} = \partial_{\mu} + i g_s T^A G_{\mu}^A $.
\par
\section{Computation of the two-loop counterterm \label{section:CHG}}
In order to determine the two-loop RGE running of $\coeff{HG}{}$, we must analyze its dependence on the renormalization scale $\mu_R$. This dependence is entirely dictated by the divergent term proportional to $1/\epsilon$ in the context of dimensional regularization, where the spacetime dimension is extended to $d=4-2\epsilon$. Consequently, our task reduces to calculating the two-loop counterterm.

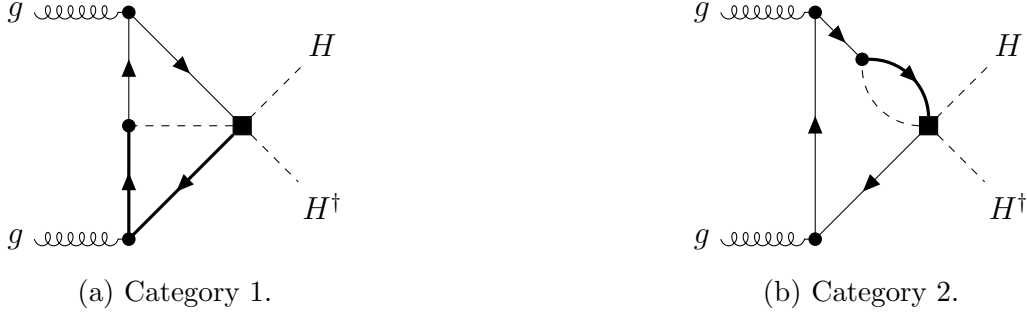
\begin{figure*}[t!]
    \centering
    \begin{subfigure}[t]{0.45\textwidth}
        \centering
        \begin{tikzpicture} 
            \begin{feynman}
                \vertex  (g1)  {$g$};
                \vertex  (gtt1) [dot, scale=\sizedot,right= of g1] {};
                \vertex (htt) [dot, scale=\sizedot,below = of gtt1] {};
                \vertex  (gtt2) [dot,scale=\sizedot,below = of htt] {};
                \vertex (ghost) [dot,scale=\sizedot,right = of htt] {};
                \vertex (C) [square dot,scale=\sizesqdot, right = of htt]  {};
                \vertex  (g2) [left=of gtt2]  {$g$};
                \vertex (h1) [above right = of C] {$H$};
                \vertex (h2) [below right = of C] {$H^\dagger$};
                \diagram* {
                    (g1)  -- [gluon] (gtt1),
                    (g2) -- [gluon] (gtt2),
                    (h1)  -- [scalar] (C) -- [scalar] (h2),
                    (gtt1) -- [fermion, line width=\lwR] (C) 
                    -- [fermion, line width=\lwL] (gtt2) --[fermion, line width=\lwL] (htt) -- [fermion, line width=\lwR] (gtt1), 
                    (C) -- [scalar] (htt)
                };
            \end{feynman}
        \end{tikzpicture}
        \caption{Category 1.}
    \end{subfigure}
    \hfill
    \begin{subfigure}[t]{0.45\textwidth}
        \centering
        \begin{tikzpicture} 
            \begin{feynman}
                \vertex  (g1)  {$g$};
                \vertex  (gtt1) [dot, scale=\sizedot,right= of g1] {};
                \vertex (htt) [dot, scale=0.1,below =  of gtt1] {};
                \vertex (htt2) [dot, scale=\sizedot,below right= 25 pt  of gtt1] {};
                \vertex  (gtt2) [dot,scale=\sizedot,below = of htt] {};
                \vertex (ghost) [dot,scale=\sizedot,right = of htt] {};
                \vertex (C) [square dot,scale=\sizesqdot, right = of htt]  {};
                \vertex  (g2) [left=of gtt2]  {$g$};
                \vertex (h1) [above right = of C] {$H$};
                \vertex (h2) [below right = of C] {$H^\dagger$};
                \diagram* {
                    (g1)  -- [gluon] (gtt1),
                    (g2) -- [gluon] (gtt2),
                    (h1)  -- [scalar] (C) -- [scalar] (h2),
                    (gtt1) -- [fermion, line width=\lwR] (htt2) -- [fermion, quarter left, line width=\lwL] (C) 
                    -- [fermion, line width=\lwR] (gtt2)
                    -- [fermion, line width=\lwR] (gtt1), 
                    (C) -- [scalar, quarter left] (htt2)
                };
            \end{feynman}
        \end{tikzpicture}
        \caption{Category 2.}    
    \end{subfigure}
    \caption{\centering Examples of diagrams contributing to the process $gg \to H^\dagger H$. The thin fermion line denotes the right-handed fields $t_R,b_R$ while the solid fermion line denotes the left-handed field $Q_L$ and the dashed line represents the massive Higgs doublet.} \label{fig:gghh1}
\end{figure*}

The computation of the counterterm is performed in the unbroken phase, where the left handed and right handed fields are independent degrees of freedom and the fermions are massless while the Higgs boson carries mass $-\mu^2$. The process that we use is $gg \to H^\dagger H$, for which a sample of the 24 diagrams (including the bottom contributions) is given in Fig.~\ref{fig:gghh1}. We have not considered tadpole diagrams because they are renormalized by the counterterms associated to the Yukawa couplings \cite{rge1}, so they do not contribute to the renormalization of $\coeff{HG}{}$. We can divide the diagrams in two categories, depending on the position of the internal Higgs propagator. 
The Feynman rules we used are reported in Appendix~\ref{sec:FeynmanRules}. 

Since a single fermion trace is involved and we are interested in the contributions to the CP even operator $\op{HG}{}$, ignoring potential effects to its CP odd counterpart $\tilde{\mathcal{O}}_{HG}$, there is no ambiguity connected to the choice of the continuation scheme for the $\gamma_5$ matrix, as we explicitly checked by comparing the result in the na\"ive dimensional regularisation scheme (NDR) \cite{CHANOWITZ1979225} and the Breitenlohner-Maison-'t Hooft-Veltman scheme (BMHV) \cite{THOOFT1972189,Breitenlohner:1977hr}.

We generate the two-loop diagrams using {\tt qgraf-3.6.5} \cite{Nogueira:1991ex}, and the manipulation to simplify the Dirac-$\gamma$ algebra are  performed with {\tt FeynCalc} \cite{Mertig:1990an, Shtabovenko:2016sxi,Shtabovenko:2020gxv}. 
The complete two loop amplitude can be written as 
\begin{equation}
i\mathcal{M}(g_A^{\mu}(k_1)g_B^{\nu}(k_2)\to H^{\dagger}H)=i g_s^2 \delta^{AB} \left[ \sum_{\psi = t,b} \left(\yuk{\psi} \frac{\coeff{\psi H}{}}{\Lambda^2}+ \yuk{\psi}^*\frac{\coeff{\psi H}{*}}{\Lambda^2} \right) \right] (k_1^{\nu}k_2^{\mu}-g^{\mu\nu}(k_1\cdot k_2)) \mathcal{A}^{(2)} ,
\end{equation}
where $\mathcal{A}^{(2)}$ corresponds to the two-loop integrand expressed as
\begin{equation}
    \mathcal{A}^{(2)} = 
    \int \prod_{i=1}^{2} \frac{d^{d}l_i}{(2 \pi)^d}\sum_{j=1}^{24}{\frac{N_j (l, k)}{\prod_{\sigma_j} D_{\sigma_j}}} .
\end{equation}
Such quantity can be extracted by means of a projection in Lorentz space, namely by contracting the amplitude with $(k_1^{\nu}k_2^{\mu}-g^{\mu\nu}(k_1\cdot k_2))$.
In the previous equation, $l$ denotes the loop momenta, $N_j (l, k)$ is the numerator containing the dot products between external and loop momenta, and $D_{\sigma_j}$ are the denominators corresponding to the $j$-th diagram.
We classify these amplitudes into two integral families (corresponding to the representative diagrams as shown in Fig.~\ref{fig:gghh1}) according to the propagator structure.
Employing \texttt{LiteRed2} \cite{Lee:2012cn,Lee:2013mka}, we perform the Integration-By-Parts (IBP) reduction to express the amplitude as a linear combination of 8 master integrals (MIs) distributed in the two families. We compute the master integrals of each family, where we do not consider the relations among the master integrals of different families. The master integrals have been analytically computed using the differential equation method~\cite{Henn:2013pwa} via Magnus exponential~\cite{Argeri:2014qva, DiVita:2014pza}, where they are expressed in terms of Harmonic polylogarithms~\cite{REMIDDI_2000, Gehrmann_2001}, around $d = 4$ space-time dimensions. The computation of the MIs have been described in detail in Appendix \ref{sec:MI}, where our results are presented. They have been also numerically cross-checked by using \texttt{AMFlow} \cite{Liu:2022chg}.

As mentioned before, the computation was performed in the unbroken phase in which the fermion lines are massless. While this simplifies the computation of the integrals, on the other hand this also requires special attention on what regards the identification of the origin of the ultraviolet (UV) divergencies, which contribute to the RGE running only. In our case, the final result of the amplitude cannot be infrared (IR) divergent as there is no corresponding real emission diagrams at the considered order in the coupling constants. Nevertheless, in intermediate steps IR divergencies can arise. Knowing they need to cancel, we need to assure that they did not cancel with UV divergencies, which for instance happens in scaleless integrals. For this reason, we have repeated our computation of the integrals numerically with \texttt{AMFlow} using a small fermion mass $m_{IR}$ as IR regulator. We find the same result as for the massless case for the pole and we assured ourselves that the finite part does not depend on $m_{IR}$ for $m_{IR}\to 0$. 

The tree-level insertion of the operator $\op{HG}{}$ reads
\begin{equation}
i\mathcal{M}(g_a^{\mu}(k_1)g_b^{\nu}(k_2)\to H^{\dagger}H)=i 4 \delta^{ab}\frac{\coeff{HG}{}}{\Lambda^2} (k_1^{\nu}k_2^{\mu}-g^{\mu\nu}(k_1\cdot k_2)).
\end{equation}
\newline
We employ the $\overline{\text{MS}}$ renormalization scheme, denoting with $\coeff{H G}{(0)}$ the bare parameter and with $\coeff{H G}{}$ the renormalized one. We set
\begin{equation}\label{eq:ctphiG1}
\coeff{H G}{} = \coeff{H G}{(0)} + \ct{H G}{}.
\end{equation}
We obtain 
\begin{equation}\label{eq:ctphiG2}
\ct{H G}{} = - \frac{3}{4 \epsilon}  g_s^2\left(\frac{1}{16 \pi^2}  \right)^2\left[ \coeff{t H}{} \yuk{t}+\coeff{t H}{*} \yuk{t}^* +\coeff{b H}{} \yuk{b}+\coeff{b H}{*} \yuk{b}^* \right].
\end{equation}
The algebraic formula connecting the counterterm with the $\beta$ function of the corresponding operator has been presented in Appendix~B of Ref.~\cite{DiNoi:2023ygk}, which, in our case, schematically reads ($\psi = t,b$): 
\begin{equation}\label{eq:ctfromRGE}
\delta \coeff{H G}{}\supset \frac{1}{ {\epsilon}} \gamma_{H G, \psi H} \coeff{\psi H}{} \frac{1}{\kappa_{H G} - \kappa_{\psi H} - 3} \;, 
\end{equation}
where 
\begin{equation}
\mu \frac{d \coeff{H G}{}}{ d \mu } \supset \gamma_{H G, \psi H} \coeff{\psi H}{}.
\end{equation} 
The $\kappa$ terms stem from the redefinitions $\coeff{X}{}\to \mu^{\kappa_X {\epsilon}} \coeff{X}{}$, performed to keep the Wilson coefficients of the dimension-six operators dimensionless. One obtains staightforwardly 
\begin{equation}
\kappa_{H G} = 2, \quad \kappa_{\psi H}=3.
\end{equation}
This allows to write, finally
\begin{equation}
\begin{aligned}
\mu \frac{d \coeff{H G}{} }{ d \mu } \supset & {\frac{3}{2}} \left(\frac{1}{16 \pi^2} \right)^2 g_s^2 \left[ \coeff{t H}{} \yuk{t}+\coeff{t H}{*} \yuk{t}^* +\coeff{b H}{} \yuk{b}+\coeff{b H}{*} \yuk{b}^* \right]  \\ 
-& 4 \left(\frac{1}{16 \pi^2} \right)^2  g_s^2  \yuk{t}\yuk{t}^{*}  \delta_{NDR} \left(\coeff{Qt}{(1)} - \frac{1}{6} \coeff{Qt}{(8)} \right) ,
\end{aligned} \label{eq:RGECHG}
\end{equation}
\begin{equation}
\delta_{NDR} =
\begin{cases}
1 & \text{(NDR)}\\
0 & \text{(BMHV)}.
\end{cases}
\label{eq:Ktg}
\end{equation}

We note that in principle also operators of the type $(\bar{\psi} \gamma_{\mu}\psi) (H^{\dagger} \overleftrightarrow{D}^\mu H)$ could give a contribution proportional to the top Yukawa coupling. Such operators will surely get contributions  similar to Fig.~\ref{fig:gghh1} (a) with a vector boson exchange. Those are proportional to the weak coupling constants and not the top Yukawa coupling. In addition, there are some  contributions that correct the top Yukawa coupling at one-loop level hence again not contributing to the running of $\coeff{HG}{}$. Finally, there can be diagrams that can be regarded as finite contributions to the operator $\coeff{tG}{}$ whose insertion into the processes $gg \to hg/hh$ is by itself divergent. Those can hence give a genuine contribution to the two-loop running of $\coeff{HG}{}$ and are $\gamma_5$ scheme dependent as shown in Ref.~\cite{DiNoi:2023ygk}.

Unless otherwise stated, we will left understood that the NDR continuation scheme is used for the two-loop contributions.

\section{Phenomenological implications} \label{section:pheno}

We test the phenomological impact of the two-loop running effect computed in this paper by studying its effect in differential distributions in the processes $pp \to hj$ and $pp \to hh$. 

We select two scenarios, differing for the values of the coefficients at the new physics scale $\Lambda$ (assuming here and in the following $\Lambda = 1 \,\text{TeV}$). The validity of this choice is tested by implementing the SMEFT fit of Ref.~\cite{ATLAS:2024lyh} in the Gaussian approximation for what concerns $\coeff{tH}{},\coeff{HG}{},\coeff{tG}{}$ and by setting the other Wilson coefficients to zero. 
The coefficients are run down to the scale $\mu_{\chi^2} = m_h$ at which we perform a $\chi^2$ test at 95 \% CL.\footnote{Since the fit considers differential measurements it would be of course interesting to check the impact of the running of the Wilson coefficients on the fit. This is though not implemented in Ref.~\cite{ATLAS:2024lyh} and remains hence beyond the scope of this work. It has though be shown in Ref.~\cite{Maltoni:2024dpn} in a toy fit that the running effect can influence a fit sizeably. } 
The four-top operators are chosen to be inside the marginalised bounds at $\order{1/\Lambda^2}$ presented in Ref.~\cite{Ethier:2021bye}. 

The numerical input we use in this paper is 
\begin{equation}
m_h = 125\, \text{GeV},\; m_t = 172.5\, \text{GeV}, \; v = 246.22\, \text{GeV},\; \alpha_s(m_Z)=0.118,\; E_{\text{coll}}= 13.6 \, \text{TeV}. 
\end{equation} 
We used the \texttt{PDF4LHC21\_40} \cite{PDF4LHCWorkingGroup:2022cjn} parton distribution functions in the \texttt{LHAPDF6} format \cite{Buckley:2014ana}.
The running effects at one loop have been implemented using \texttt{RGESolver} \cite{DiNoi:2022ejg} \texttt{V1.0.1}. The two-loop effect have been included by modifying the code.

Finally, we emphasise that we include only in $\coeff{HG}{}$ two-loop running effects. In the other operators, entering only at loop level, those would contribute to an order higher.

\begin{figure}[h!]
    \centering
    \begin{subfigure}[t]{0.48\textwidth}
        \centering
        \includegraphics[height=5cm]{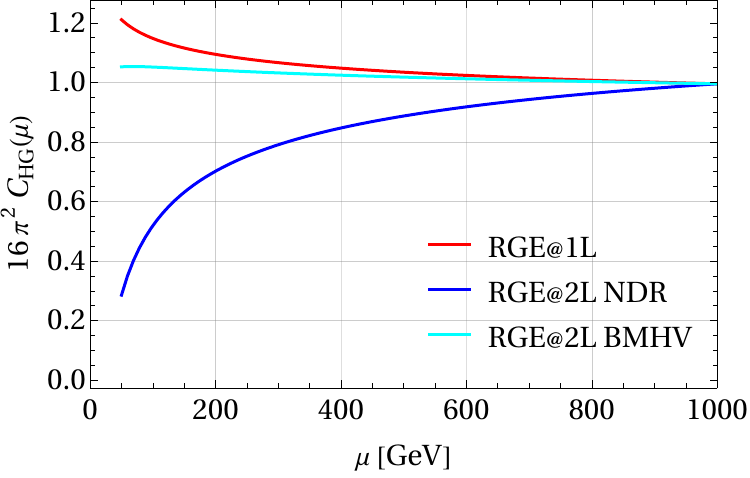}
    \end{subfigure}
    \hfill
    \begin{subfigure}[t]{0.48\textwidth}
        \centering
        \includegraphics[height=5cm]{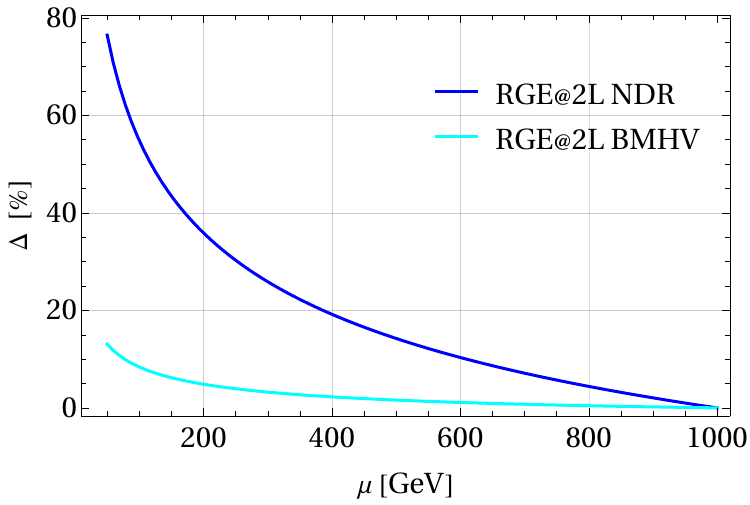}
    \end{subfigure}
    \caption{\centering Comparison between one and two loop running of the Wilson coefficient $\coeff{HG}{}$ in S1 as a function of the renormalization scale $\mu$. \textit{Left:} Value of $\coeff{HG}{}$. \textit{Right:} Percentual difference between one-loop and two-loop RGE running computed as $\Delta = \left(\coeff{HG}{\text{1L}}(\mu)-\coeff{HG}{\text{2L}}(\mu) \right)/\coeff{HG}{\text{1L}}(\mu)$. }
   
    \label{fig:CHGS1}
\end{figure}

The first set-up for the Wilson coefficients (scenario 1) follows the loop counting discussed in Ref.~\cite{Buchalla:2022vjp}:
\begin{equation} \label{eq:initialconditionsS1}
\text{S1}: \quad \coeff{tH}{}(\Lambda) = 1, \; \coeff{HG}{}(\Lambda) = \frac{1}{16 \pi^2}, \; \coeff{tG}{}(\Lambda) = - \frac{1}{16 \pi^2}, \; \coeff{Qt(1,8)}{}(\Lambda) = -10, \; \coeff{H}{}(\Lambda)=0. 
\end{equation}
Several reasons suggest that new physics couples dominantly with the third generation, generating large Wilson coefficients for the SMEFT operators involving such fields. A complete overview of the tree-level matching of NP models can be found in Ref.~\cite{deBlas:2017xtg}. However, we mention the extensions featuring the new scalars $\Phi \sim (\mathbf{8},\mathbf{2})_{1/2}$ or~$\varphi\sim (\mathbf{1},\mathbf{2})_{1/2}$, denoting the $\text{SU(3)}_{\text{C}} \otimes \text{SU(2)}_{\text{L}}\otimes\text{U(1)}_{{\mathsf{y}}}$ quantum numbers as $({\mathbf{R}}_{\text{C}},{\mathbf{R}}_{\text{L}})_{\mathsf{y}}$, have been discussed in Ref.~\cite{DiNoi:2023ygk} in NDR and BMHV and the vector boson $\mathcal{Y}_{5} \sim (\bar{6},2)_{-5/6}$ generates only $\coeff{Qt(1,8)}{}$ at tree level. 
This case serves as a showcase of a scenario in which the two-loop running effects can be important. It should be stressed that this scenario is chosen in such a way the four-top operators and $\op{tH}{}$ contribute with the same sign in the RGE of $\op{HG}{}$, enhancing the effect. \footnote{Flipping, for instance, the sign of the four-top operator leads to a little bit smaller effects maximally amounting to $-14\%$ at low transverse momenta. } Moreover, we let the four-top operators be larger since they are less constrained \cite{Ethier:2021bye}. 
Within this set-up the two loop contributions in the NDR scheme (blue line) affect the running of $\coeff{HG}{}$ significantly (up to $80 \%$), as shown in Fig.~\ref{fig:CHGS1}, if compared to the one loop distribution (red line). The effects are smaller in the BMHV scheme (cyan line), where the four-top contribution in the RGE of $\coeff{HG}{}$ vanishes according to Eq.~\eqref{eq:RGECHG}.

\begin{figure}[h!]
    \centering
        \includegraphics[height=5cm]{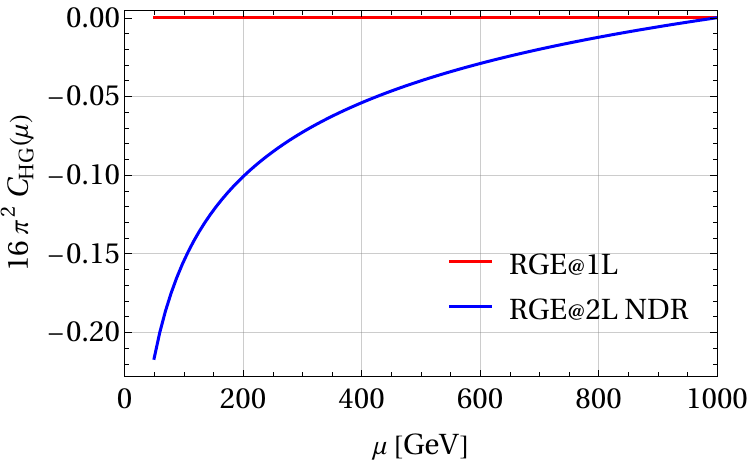}
        \caption{\centering $\coeff{HG}{}(\mu)$ as a function of the renormalization scale $\mu$ at one-loop (red line) and two-loop (blue) level in S2.}
        \label{fig:CHGRunningS2}
\end{figure}

As a second scenario, we study the case in which only the operator $\coeff{tH}{}$ is non-vanishing at the high-energy scale:
\begin{equation} \label{eq:initialconditionsS2}
\text{S2}: \quad \coeff{tH}{}(\Lambda) = 3, \; \coeff{HG}{}(\Lambda) = 0, \; \coeff{tG}{}(\Lambda) =0,\;\coeff{Qt(1,8)}{}(\Lambda) =0, \; \coeff{H}{}(\Lambda)=0. 
\end{equation}
This scenario allows to focus on the term which was computed for the first time in this work.
The evolution of $\coeff{HG}{}$ as a function of the renormalization scale is displayed in Fig.~\ref{fig:CHGRunningS2} for scenario 2.

\subsection{Higgs $p_{T,h}$-spectrum \label{section:Hj}}
We show the impact on the transverse momentum distribution in $pp \to hj$ using the results in Ref.~\cite{Grazzini:2018eyk}, where the results for diagrams with a single insertion of dimension six SMEFT operators have been presented for the first time. This means that the matrix element is computed at $\order{1/\Lambda^2}$ and the cross section at $\order{1/\Lambda^4}$ (but no dimension eight operators are taken into account). 
For the numerical evaluation of the cross section we use an in-house code in \texttt{Mathematica} and \texttt{C++}. We employ \texttt{RunDec} \cite{Chetyrkin:2000yt,Herren:2017osy} to compute the running of $\alpha_s$ and \texttt{ManeParse} \cite{Clark:2016jgm} to access the PDF. 
Apart from when otherwise stated, we use a dynamical renormalization and factorization scale 
\begin{equation} \label{eq:dynscale}
\mu_R = \mu_F = \frac{1}{2} \sqrt{m_h^2 + p_{T,h}^2}.
\end{equation}

\begin{figure}[h!]
    \centering
    \begin{subfigure}[t]{0.48\textwidth}
        \centering
        \includegraphics[height=5cm]{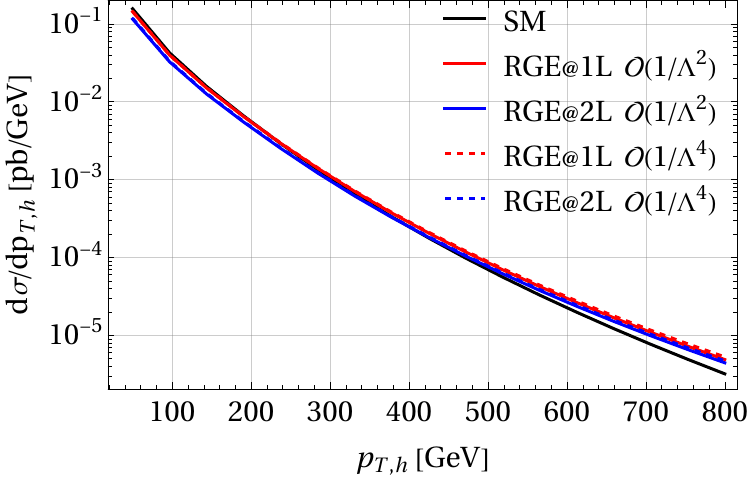}
        \end{subfigure}
    \hfill
   \begin{subfigure}[t]{0.48\textwidth}
        \centering
        \includegraphics[height=5.07cm]{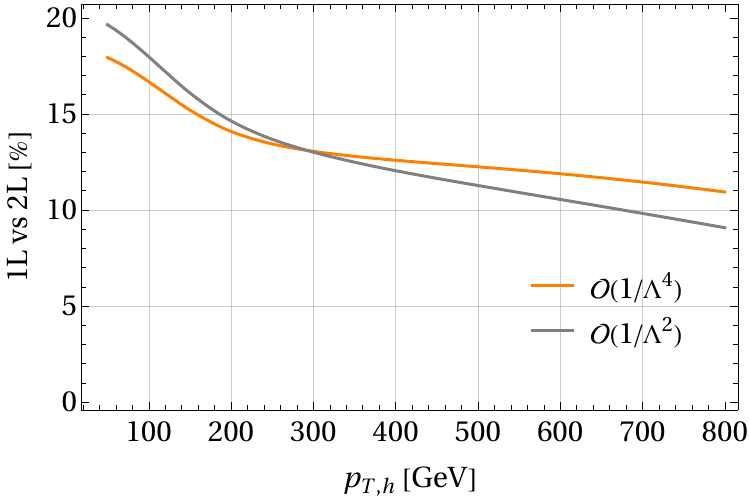}
        \end{subfigure}
\caption{\centering Comparison between one-loop (red line) and two-loop (blue line) running of the Wilson coefficient $\coeff{HG}{}$ in the transverse momentum distribution in S1. \textit{Left:} Differential distribution in $p_{T,h}$ in the SM and in the SMEFT. \textit{Right:} Percentual difference between one-loop (1L) and two-loop (2L) running defined as (1L-2L)/1L.  }
    \label{fig:hjS1}
\end{figure}

\begin{figure}[h!]
        \centering
        \includegraphics[height=5.07cm]{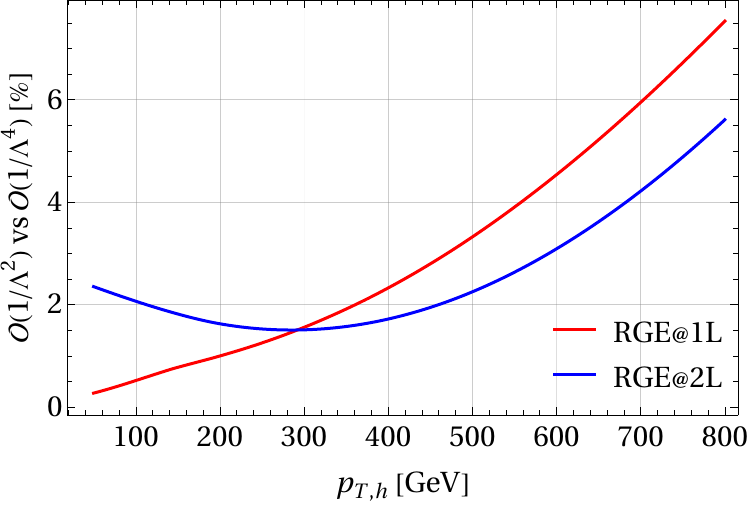}
        \caption{\centering Percentual difference between linear (lin) and quadratic (quad) order in $1/\Lambda^2$ defined as (quad-lin)/quad in S1.}
    \label{fig:hjS1difference}
\end{figure}

For scenario 1, we show the cross section as a function of the transverse momentum of the Higgs boson, $p_{T,h}$, in Fig.~\ref{fig:hjS1}. The SM distribution (black line) is shown for comparison. Within this set-up, the two-loop contributions have a sizeable impact, compatible with Fig.~\ref{fig:CHGS1}. The difference between the two cases ranges in the interval $ [10, 20] \, \%$ in the considered transverse momentum interval. We note that the inclusion of the two-loop terms is much more important as in $t\bar{t}h$ \cite{DiNoi:2023onw}, which can be easily explained by the fact that there the Wilson coefficient $\coeff{HG}{}$ enters at the same order than the SM contribution, namely at tree level, contrary to Higgs+jet.  Moreover, the difference stemming from the inclusion of quadratic terms in the $1/\Lambda^2$ expansion increases the cross section in the high $p_{T,h}$ region by less than $10 \, \%$. Including such terms does not change significantly the difference in the distribution between the one and two-loop running, shown in Fig.~\ref{fig:hjS1difference}. We conclude that the difference between one- and two-loop running for this scenario is bigger than the difference between $\mathcal{O}(1/\Lambda^2)$ and $\mathcal{O}(1/\Lambda^4)$ which could be regarded as a proxy of the EFT uncertainty.

\begin{figure}[h!]
    \centering
    \begin{subfigure}[t]{0.48\textwidth}
        \centering
        \includegraphics[height=5cm]{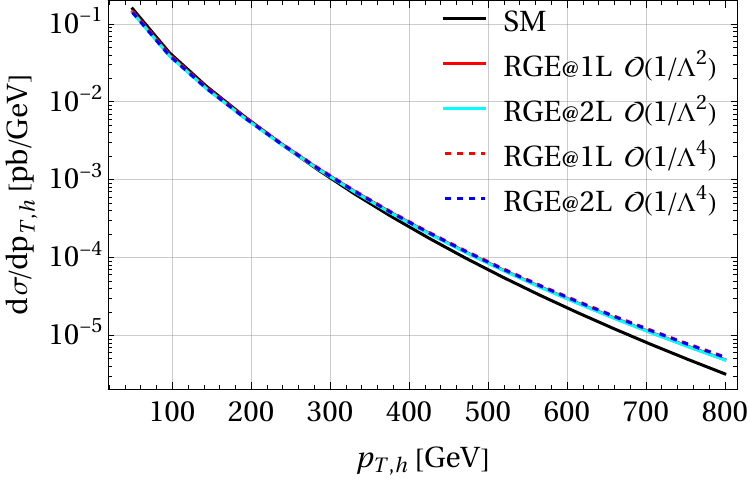}
        \end{subfigure}
    \hfill
   \begin{subfigure}[t]{0.48\textwidth}
        \centering
        \includegraphics[height=5.07cm]{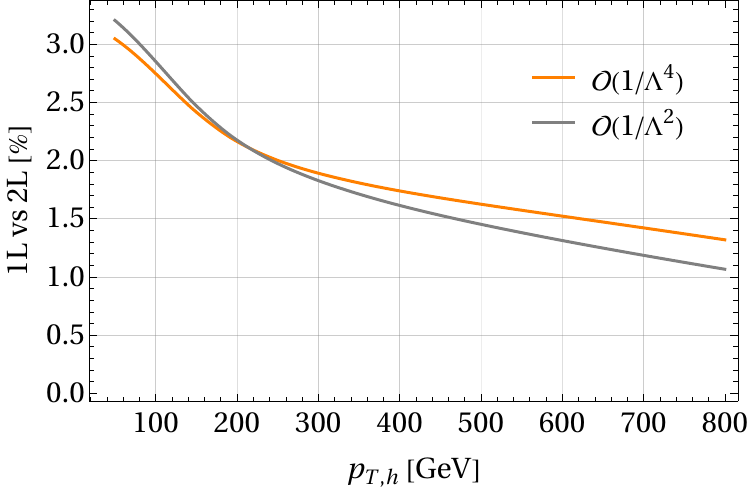}
        \end{subfigure}
\caption{\centering Comparison between one-loop (red line) and two-loop (blue line) running of the Wilson coefficient $\coeff{HG}{}$ in the transverse momentum distribution in S1 in the BMHV scheme. \textit{Left:} Differential distribution in $p_{T,h}$ in the SM and in the SMEFT. \textit{Right:} Percentual difference between one-loop (1L) and two-loop (2L) running defined as (1L-2L)/1L.  }
    \label{fig:hjS1BMHV}
\end{figure}

We repeat the analysis employing the BMHV scheme, presenting the results in Fig.~\ref{fig:hjS1BMHV}. We observe that the difference between the one- and two-loop running is reduced with respect to the NDR case in Fig.~\ref{fig:hjS1}. This can be understood from Eq.~\eqref{eq:RGECHG} and Fig.~\ref{fig:CHGS1}: in BMHV, the largely unconstrained four-top operators do not contribute to the running of $\coeff{HG}{}$.

\begin{figure}[h]
    \centering
    \begin{subfigure}[t]{0.48\textwidth}
        \centering
        \includegraphics[height=5cm]{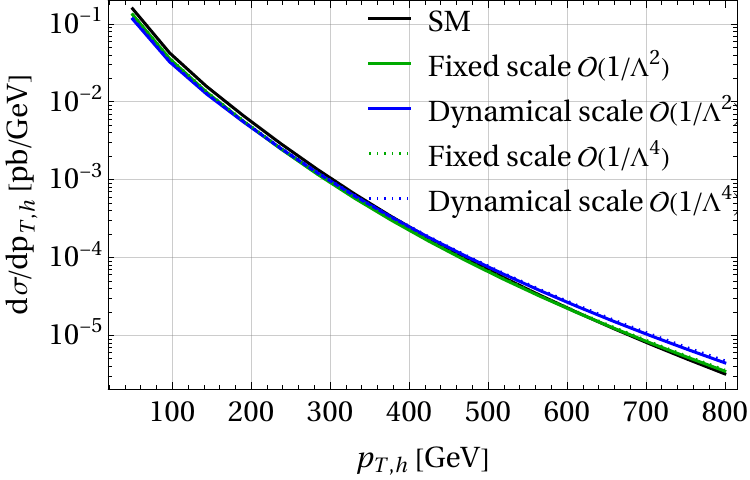}
        \end{subfigure}
    \hfill
    \begin{subfigure}[t]{0.48\textwidth}
        \centering
        \includegraphics[height=5cm]{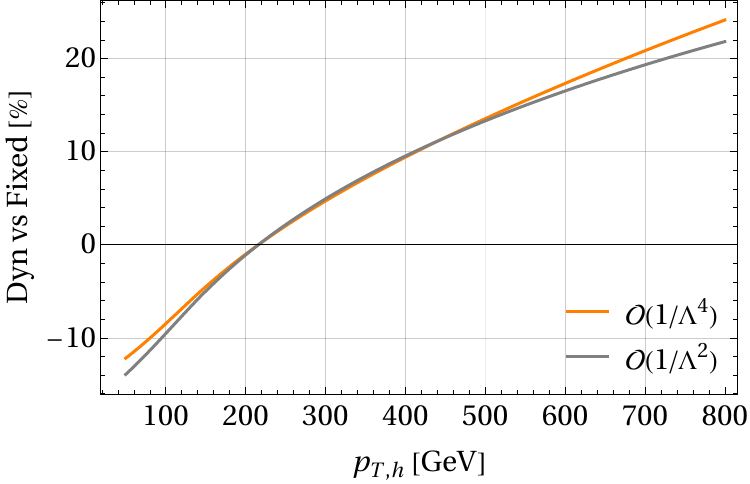}
        \end{subfigure}
\caption{\centering Comparison between a dynamical renormalization scale $\mu_R = \frac{1}{2} \sqrt{m_h^2+p_{T,h}^2}$ for the Wilson coefficients and a fixed one $\mu_R = m_h$ in
the transverse momentum distribution in S1. \textit{Left}: Differential distribution in $p_{T,h}$
in the SM and in the SMEFT. \textit{Right}: Percentual difference defined as (Dyn-Fix)/Dyn.}    \label{fig:hjS1FixedVSDyn}
\end{figure}

We also present a comparison between two different choices of renormalization scales for the Wilson coefficients, namely a fixed renormalization scale $\mu_R = m_h$ and the dynamical scale of Eq.~\eqref{eq:dynscale} in Fig.~\ref{fig:hjS1FixedVSDyn}. It should be remarked that we choose a fixed renormalization scale only for the Wilson coefficients, since we still employ the dynamical scale in Eq.~\eqref{eq:dynscale} as a factorization scale and as renormalization scale for $\alpha_s$. This choice, despite being potentially incoherent since different renormalization scales are used in the computation, allows to highlight the running effects of the Wilson coefficients alone. Within this scenario, employing a fixed scale renormalization scale leads to an overestimation (about $15\,\%$) at low $p_{T,h}$ and an underestimation at high $p_{T,h}$ (about 25 \%).
The two curves cross at $p_{T,h} \approx 215 \, \text{GeV}$, where the two scales coincide. 

Figure~\ref{fig:hjS1FixedVSDyn} clearly emphasises again the importance of including RGE running effects in the Higgs $p_{T,h}$ distribution. 
\par
\begin{figure}[h!]
    \centering
    \begin{subfigure}[t]{0.48\textwidth}
        \centering
        \includegraphics[height=5cm]{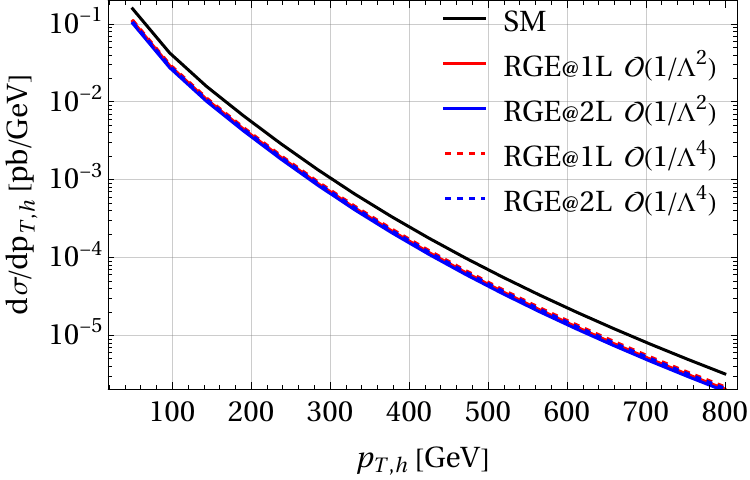}
        \end{subfigure}
    \hfill
   \begin{subfigure}[t]{0.48\textwidth}
        \centering
        \includegraphics[height=5.07cm]{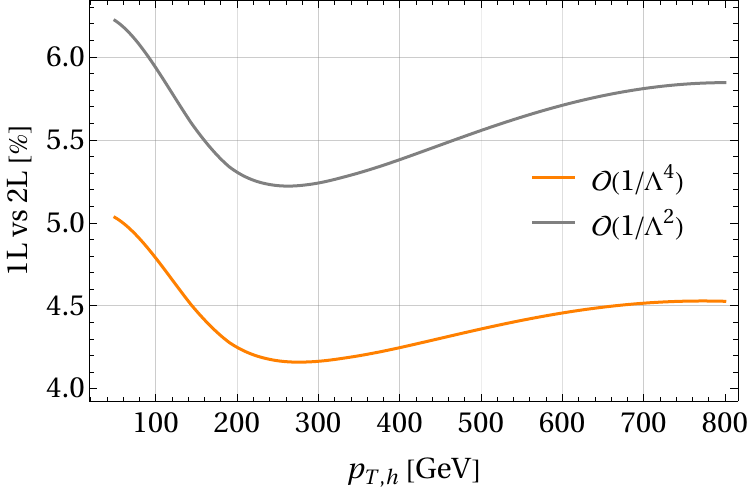}
        \end{subfigure}
\caption{\centering Comparison between one-loop (red line) and two-loop (blue line) running of the Wilson coefficient $\coeff{HG}{}$ in the transverse momentum distribution in S2. \textit{Left:} Differential distribution in $p_{T,h}$ in the SM and in the SMEFT. \textit{Right:} Percentual difference between one-loop (1L) and two-loop (2L) running defined as (1L-2L)/1L.  }
    \label{fig:hjS2}
\end{figure}

In scenario 2, shown in Fig.~\ref{fig:hjS2} we notice a larger deviation with respect to the SM at low momentum, leading to a smaller cross section due to the negative sign of $\coeff{HG}{}$. Including the quadratic terms slightly increases the cross section, as expected. The difference between the one-loop running and the two-loop running is a bit smaller than in scenario 1 but still of the order of {5 \%}. 

\subsection{Higgs Pair Production \label{section:HH}}
We now discuss Higgs pair production via gluon fusion which in the SM is mediated by triangle and box diagrams of top quarks, whereas in the SMEFT it gets modified by the operators discussed above. In particular, also here the Wilson coefficient $\coeff{HG}{}$ enters via tree-level Feynman diagrams. 
\par
For the process, usually a dynamical scale choice of $\mu_R=m_{hh}/2$ is adopted \cite{deFlorian:2015moa, DiMicco:2019ngk}, where $m_{hh}$ denotes the invariant mass of the Higgs boson pair. 
In order to show the effect of the RGE running we use a private version of \texttt{hpair} \cite{hpair, Dawson:1998py}  based on \cite{Grober:2015cwa} translated to the Warsaw basis to compute the LO differential cross section. In addition, we have implemented the chromomagnetic operator using the results of \cite{Heinrich:2023rsd}. 
\par
We include in the analysis terms up to $\mathcal{O}(1/\Lambda^2)$ but refer to \cite{Heinrich:2022idm, Alasfar:2023xpc} for a discussion of the inclusion of additional terms of $\mathcal{O}(1/\Lambda^4)$ at dimension six in the matrix element squared. We mention though that the difference between $\mathcal{O}(1/\Lambda^2)$ and $\mathcal{O}(1/\Lambda^4)$ for $\Lambda=1\text{ TeV}$ and Wilson coefficients of order 1 can be quite large due to the delicate interference structure of box and triangle diagrams in the process.
\par
\begin{figure}[h!]
    \centering
    \begin{subfigure}[t]{0.48\textwidth}
        \centering
        \includegraphics[height=5cm]{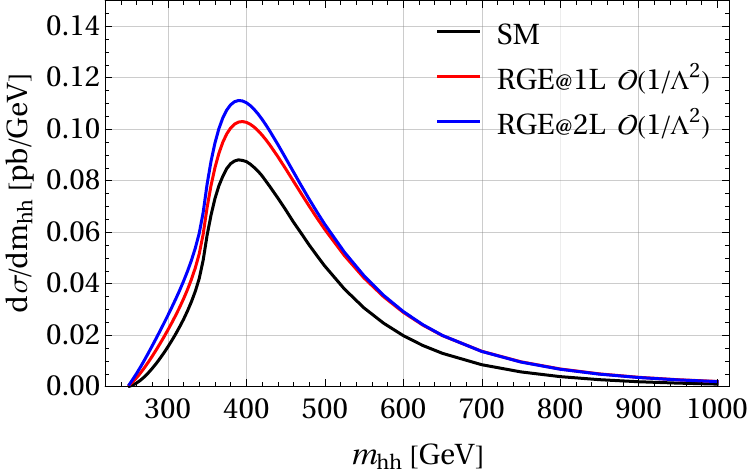}
        \end{subfigure}
    \hfill
   \begin{subfigure}[t]{0.48\textwidth}
        \centering
        \includegraphics[height=5.07cm]{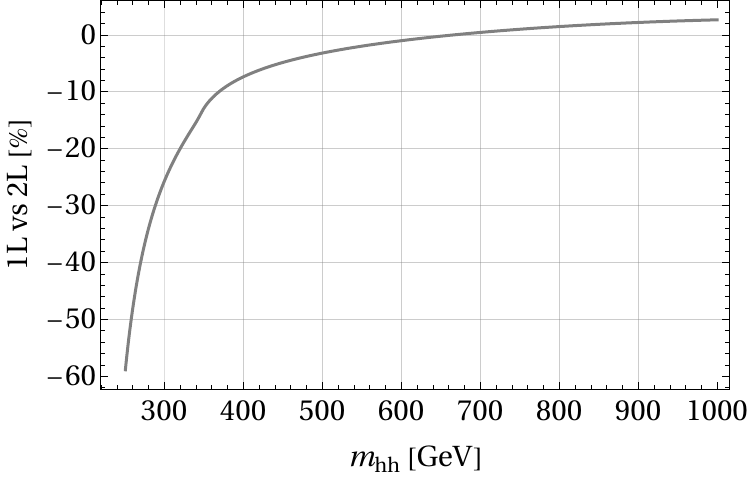}
        \end{subfigure}
\caption{\centering Comparison between one-loop and two-loop running of the Wilson coefficient $\coeff{HG}{}$ in the invariant mass distribution in S1. \textit{Left:} Differential distribution in $m_{hh}$ in the SM and in the SMEFT. \textit{Right:} Percentual difference between one-loop (1L) and two-loop (2L) running results defined as (1L-2L)/1L. }
    \label{fig:hhS1}
\end{figure}
In Fig.~\ref{fig:hhS1} we show the invariant mass distribution of the LO cross section using the RGE running including only one-loop terms (red line) and using the RGE running including also two-loop terms (blue line) and compare it to the SM distribution (black line). While {in S1} the difference between the one-loop and two-loop RGE running is quite large for small $m_{hh}$, for $m_{hh}> 400\text{ GeV}$ it remains below 10\%. This can be explained by the fact that the cross section is very small close by the threshold so little changes can make a big effect in the ratio. In addition, for small $m_{hh}$ there is a quite precise interference between box and triangle diagrams that renders the SM cross section very small. This gets spoilt by the SMEFT operators and their running. Finally, one should also note that by construction the running effects at large $m_{hh}$ are smaller since we run the coefficients starting from the high scale $\Lambda=1\text{ TeV}$.

\begin{figure}[h!]
    \centering
    \begin{subfigure}[t]{0.48\textwidth}
        \centering
        \includegraphics[height=5cm]{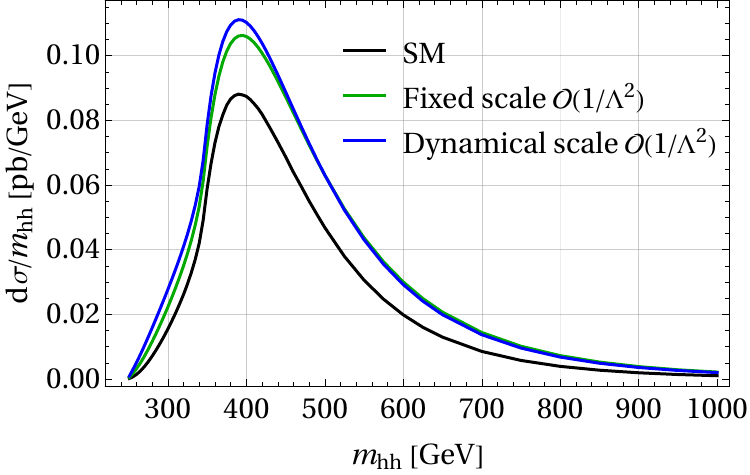}
\end{subfigure}
    \hfill
    \begin{subfigure}[t]{0.48\textwidth}
        \centering
        \includegraphics[height=5cm]{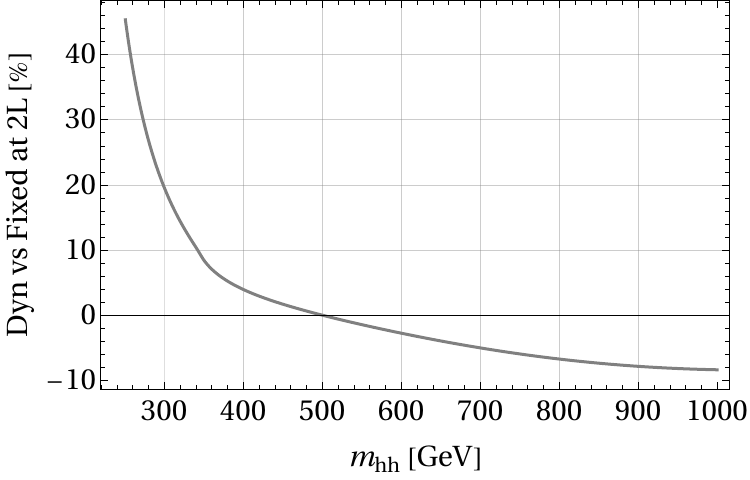}
        \end{subfigure}
  \caption{Comparison between a dynamical renormalization scale $\mu_R = m_{hh}/2$ for the Wilson coefficients and a fixed one $\mu_R = 2 \,m_h$ in
the invariant mass distribution in S1. \textit{Left}: Differential distribution in $m_{hh}$
in the SM and in the SMEFT. \textit{Right}: Percentual difference defined as (Dyn-Fix)/Dyn.} \label{fig:hhS1FixedVSDyn}
\end{figure}

In Fig.~\ref{fig:hhS1FixedVSDyn} we show the impact of the choice of a dynamical renormalization scale with respect to a fixed one, chosing in the latter case $\mu_R=2\, m_h$ for the Wilson coefficients only, while keeping $\alpha_s$ a running parameter. Indeed, also here one sees that the largest effect comes from closeby the threshold while being up to -10\% at large $m_{hh}$. Obviously, given our choice of the fixed renormalization scale at $m_{hh}=500 \text{ GeV}$ the difference is 0. Still, the plot clearly shows the importance of including the running of the Wilson coefficients over the invariant Higgs mass distribution in Higgs pair production. 
\par
\begin{figure}[h!]
    \centering
    \begin{subfigure}[t]{0.48\textwidth}
        \centering
        \includegraphics[height=5cm]{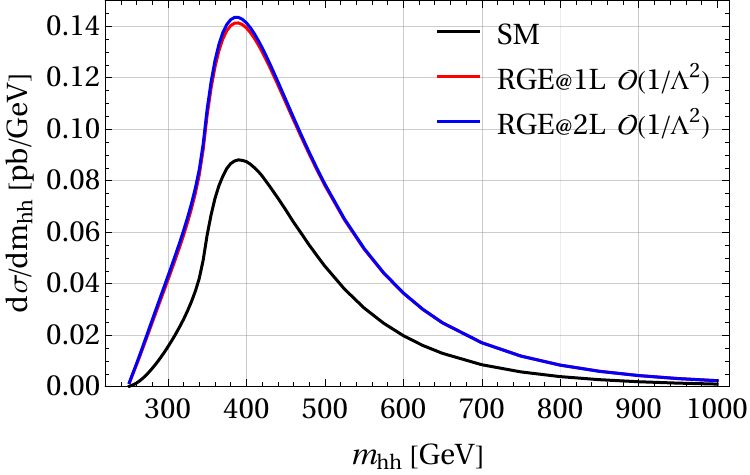}
        \end{subfigure}
    \hfill
   \begin{subfigure}[t]{0.48\textwidth}
        \centering
        \includegraphics[height=5.07cm]{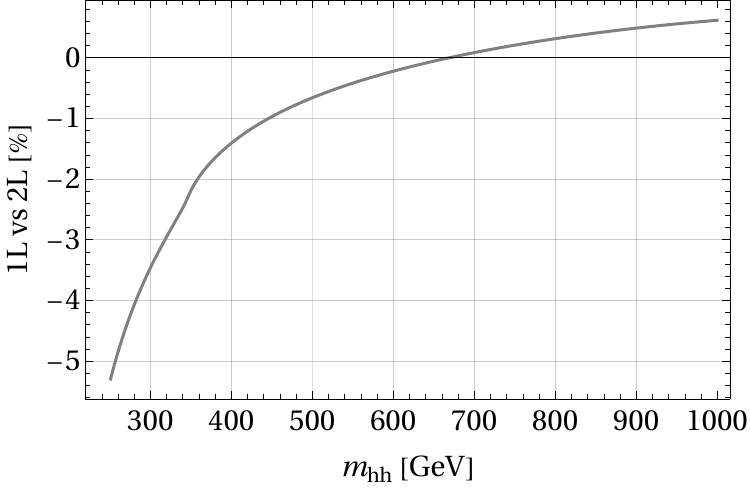}
        \end{subfigure}
\caption{\centering Comparison between one-loop (1L) and two-loop (2L) running of the Wilson coefficient $\coeff{HG}{}$ in the invariant mass distribution in S2. \textit{Left:} Differential distribution in $m_{hh}$ in the SM and in the SMEFT. \textit{Right:} Percentual difference between one-loop (1L) and two-loop (2L) running results defined as (1L-2L)/1L. }
    \label{fig:hhS2}
\end{figure}

In Fig.~\ref{fig:hhS2} we show the results for scenario 2. In this case the effects are much smaller. Indeed, there is little difference between one- and two-loop RGE running apart from again closeby the threshold where the effect is up to {5 \%}.  

\section{Conclusion \label{section:conclusion}}
We have studied the RGE running of SMEFT at two-loop order including for the first time the two-loop terms proportional to $\coeff{tH}{}$ into the running of $\coeff{HG}{}$. In Higgs physics, in various processes such as single Higgs production in gluon fusion, Higgs+jet or Higgs pair production the operator $\op{HG}{}$ enters at tree level while typically other operators, as well as the SM contribution, enter at one-loop level. For this reason, a leading RGE analysis of the SMEFT operators should contain the one-loop running of the operators that enter at one loop level (as the SM contribution) and the two-loop running of $\coeff{HG}{}$. 
At one-loop level $\coeff{HG}{}$ gets renormalised by itself and $\coeff{tG}{}$. 
Adopting a loop counting of $\coeff{HG}{}$ and $\coeff{tG}{}$ in weakly-interacting and renormalizable models leads to the conclusion that they can be generated only at one-loop level, which would lead to a two-loop effect in the running.
This implies that potentially tree-level generated operators mixing at the two-loop level with $\coeff{HG}{}$ should be consistently taken into account adopting the aforementioned loop counting. 
Given the result of our computation, we could include the RGE running effects in  Higgs+jet production and the dominant Higgs pair production process via gluon fusion considering also four-fermion operators and $\mathcal{O}_{tH}$.

We studied the phenomenological effect of our contribution in the NDR scheme and found that for the scenarios we considered the difference between one and two-loop running effects can be up to 20\%. For Higgs pair production, they seem typically smaller but for closeby the threshold, where they are even larger than 20\%. In general, it makes an important numerical difference to include those contributions. This motivates to compute the complete two-loop RGEs at least for operators such as $\coeff{HG}{}$ that can enter a loop-order lower than the SM contributions in phenomenologically important processes. 

When the BMHV scheme is employed, we observe a milder effect with respect to the NDR scheme due to the absence of the four-top contribution in the RGE of $\coeff{HG}{}$, as evident from Eq.~\eqref{eq:RGECHG}. This study showcases once again how the Wilson coefficients need to be interpreted coherently with the employed continuation scheme, as discussed in Ref.~\cite{DiNoi:2023ygk}.

Moreover we have shown that using a dynamical renormalization scale for the Wilson coefficients, compared to a fixed scale choice, has sizeable effects.

Finally, we note that we have performed our analysis only at the LO level. Since next-to-leading order (NLO) QCD corrections are both sizeable for the Higgs $p_{T,h}$ distribution \cite{Jones:2018hbb, Chen:2021azt, Bonciani:2022jmb} and Higgs pair production \cite{Borowka:2016ehy, Borowka:2016ypz, Baglio:2018lrj, Baglio:2020ini}, it would be interesting to study the running effects of the SMEFT operators for the NLO QCD corrected differential distributions. The RGE running effects should then also be included into the available Monte Carlo tools, e.g. for Higgs pair production \cite{Heinrich:2017kxx, Heinrich:2019bkc, Heinrich:2020ckp, Heinrich:2022idm, Bagnaschi:2023rbx}. We note that in particular the running of the top quark mass within the SM already leads to a sizeable uncertainty on the Higgs pair production cross section \cite{Baglio:2020wgt} and the interplay in the SMEFT with the running of the Wilson coefficients might be interesting to study. Our results hint also to the fact that electroweak corrections in SMEFT might be much more sizeable than for the SM (see e.g. \cite{Bi:2023bnq, Heinrich:2024dnz} for the electroweak corrections to Higgs pair production in full mass dependence). 

Given the numerical importance, it would be interesting to compute the full RGE at two-loop order for the effective Higgs gluon coupling. The procedure we have established in this paper and in \cite{DiNoi:2023ygk} can be applied straightforwardly to the missing pieces.

\section*{Note Added}
During the computation of the two-loop contributions of the operators of class 3 and 7 to the RGE of the Higgs-gluon coupling in the SMEFT in Ref.~\cite{DiNoi:2025tka} we noticed a mistake in our procedure, which let to an additional factor of 1/2 in Eq.~\eqref{eq:RGECHG} which we have corrected in this version. 
The effect in the figures of our phenomenological studies is small when also the four-top operators are included (i.e.~all plots referring to S1 in the NDR scheme) while when they are absent (i.e.~S1 in the BMHV scheme and S2) the net effect of two-loop vs. one-loop RGE effects is halved with respect to the previous versions.
\section*{Acknowledgements}
We thank Elisa Balzani and Marco Vitti for their contributions in the early stages of this project and Gudrun Heinrich, Jannis Lang, Pierpaolo Mastrolia and Marco Vitti for comments on the manuscript. 
Moreover, we thank Emanuele Bagnaschi, Giulio Crisanti, Stefano Laporta, Pierpaolo Mastrolia, Matteo Pegorin, Michael Spira and Simone Tentori for useful discussions and Konstantin Schmid for help sorting out the right factors for the  chromomagnetic operator contribution to Higgs pair production. 
The work of RG and MKM is supported in part by the Italian MUR Departments of Excellence grant 2023-2027 ”Quantum Frontiers” and the ICSC – Centro Nazionale di Ricerca in High Performance Computing, Big Data and Quantum Computing, funded by European Union – NextGenerationEU. 
The work of RG is supported by the University of Padua under the 2023 STARS Grants@Unipd programme (Acronym and title of the project: HiggsPairs -- Precise Theoretical Predictions for Higgs pair production at the LHC). 
The authors acknowledge support from the COMETA COST Action CA22130 and the Iniziativa Specifica ``Physics at the Energy, Intensity, and Astroparticle Frontiers'' (APINE) of the Istituto Nazionale di Fisica Nucleare (INFN). 
The Feynman diagrams shown in this work
were drawn with \texttt{TikZ-Feynman} \cite{Ellis:2016jkw}.
\appendix
\section{Feynman Rules} \label{sec:FeynmanRules}
We report in this Section the Feynman rules we employed in our computation. We use $m_j$ ($A_j$) to denote an index in the fundamental (adjoint) representation of $SU(3)_{\text{C}}$ and $i_j$ to denote an index in the fundamental representation of $SU(2)_{\text{W}}$. The numbers refer to the particles in the figure.
The SM Feynman rules are:
\begin{subequations}
\begin{alignat}{4}\label{FR:yu}
 \begin{tikzpicture}[baseline=(gqq)]
            \begin{feynman}[small]
                \vertex  (g)  {$g^1$};
                \vertex  (gqq) [dot, scale=\sizedot,right= of g] {};
                \vertex (q1) [above right =  of gqq] {$Q_L^2$};
                \vertex (q2) [below right =  of gqq] {$Q_L^3$};
                \diagram* {
                    (g)  -- [gluon] (gqq),
                    (q1) -- [fermion, line width=\lwL] (gqq) -- [fermion, line width=\lwL] (q2)
                };
            \end{feynman}
        \end{tikzpicture} &=-i g_s \gamma^{\mu_1 } T^{A_1}_{m_3m_2} \id{i_3 i_2},\quad
        \begin{tikzpicture}[baseline=(gqq)]
            \begin{feynman}[small]
                \vertex  (g)  {$g^1$};
                \vertex  (gqq) [dot, scale=\sizedot,right= of g] {};
                \vertex (q1) [above right =  of gqq] {$t_R^2/b_R^2$};
                \vertex (q2) [below right =  of gqq] {$t_R^3/b_R^3$};;
                \diagram* {
                    (g)  -- [gluon] (gqq),
                    (q1) -- [fermion, line width=\lwR] (gqq) -- [fermion, line width=\lwR] (q2)
                };
            \end{feynman}
        \end{tikzpicture} =-i g_s \gamma^{\mu_1 } T^{A_1}_{m_3m_2},
        \\        
 \begin{tikzpicture}[baseline=(hqq)]
            \begin{feynman}[small]
                \vertex  (h)  {$H^{1}$};
                \vertex  (hqq) [dot, scale=\sizedot,right= of h] {};
                \vertex (q1) [above right =  of hqq] {$Q^2_L$};
                \vertex (q2) [below right =  of hqq] {$t^3_R$};
                \diagram* {
                    (h)  -- [scalar] (hqq),
                    (q1) -- [fermion, line width=\lwL] (hqq) -- [fermion, line width=\lwR] (q2)
                };
            \end{feynman}
        \end{tikzpicture} &= + i \varepsilon_{i_1 i_2} \id{m_3 m_2} \yuk{t} , \qquad \begin{tikzpicture}[baseline=(hqq)]
            \begin{feynman}[small]
                \vertex  (h)  {$H^{1,\dagger}$};
                \vertex  (hqq) [dot, scale=\sizedot,right= of h] {};
                \vertex (q1) [above right =  of hqq] {$t_R^2$};
                \vertex (q2) [below right =  of hqq] {$Q_L^3$};
                \diagram* {
                    (h)  -- [scalar] (hqq),
                    (q1) -- [fermion, line width=\lwR] (hqq) -- [fermion, line width=\lwL] (q2)
                };
            \end{feynman}
        \end{tikzpicture} =  -i \varepsilon_{i_3 i_1} \id{m_3 m_2} \yuk{t}^* ,
        \\
  \begin{tikzpicture}[baseline=(hqq)]
            \begin{feynman}[small]
                \vertex  (h)  {$H^{1,\dagger}$};
                \vertex  (hqq) [dot, scale=\sizedot,right= of h] {};
                \vertex (q1) [above right =  of hqq] {$Q^2_L$};
                \vertex (q2) [below right =  of hqq] {$b^3_R$};
                \diagram* {
                    (h)  -- [scalar] (hqq),
                    (q1) -- [fermion, line width=\lwL] (hqq) -- [fermion, line width=\lwR] (q2)
                };
            \end{feynman}
        \end{tikzpicture} &=  -i \id{i_1 i_2} \id{m_3 m_2} \yuk{b}, \qquad \begin{tikzpicture}[baseline=(hqq)]
            \begin{feynman}[small]
                \vertex  (h)  {$H^{1}$};
                \vertex  (hqq) [dot, scale=\sizedot,right= of h] {};
                \vertex (q1) [above right =  of hqq] {$b_R^2$};
                \vertex (q2) [below right =  of hqq] {$Q_L^3$};
                \diagram* {
                    (h)  -- [scalar] (hqq),
                    (q1) -- [fermion, line width=\lwR] (hqq) -- [fermion, line width=\lwL] (q2)
                };
            \end{feynman}
        \end{tikzpicture} =  -i \id{i_3 i_1} \id{m_3 m_2} \yuk{b}^*.
\end{alignat}
\end{subequations}
The operators $\op{tH}{},\op{b H}{}$ generate the following interaction vertices:

\begin{subequations}
\begin{align}
\begin{tikzpicture}[baseline=(hqq)]
            \begin{feynman}[small]
                \vertex  (h)  {$H^{2,\dagger}$};
                \vertex  (hqq) [dot, scale=\sizedot,right= of h] {};
                \vertex  (h2) [above left = of hqq] {$H^1$};
                \vertex  (h3) [below left = of hqq] {$H^{3,\dagger}$};
                \vertex (q1) [above right =  of hqq] {$t^4_R$};
                \vertex (q2) [below right =  of hqq] {$Q^5_L$};
                \diagram* {
                    (h)  -- [scalar] (hqq),
                    (q1) -- [fermion, line width=\lwR] (hqq) -- [fermion, line width=\lwL] (q2),
                    (h2) -- [scalar] (hqq) --[scalar] (h3),
                };
            \end{feynman}
        \end{tikzpicture} &= + i\; \coeff{t H }{}\left[ \id{i_1 i_2} \varepsilon_{i_5 i_3} + \id{i_1 i_3} \varepsilon_{i_5 i_2} \right] \id{m_5 m_4}  , \\
        \begin{tikzpicture}[baseline=(hqq)]
             \begin{feynman}[small]
                \vertex  (h)  {$H^{2,\dagger}$};
                \vertex  (hqq) [dot, scale=\sizedot,right= of h] {};
                \vertex  (h2) [above left = of hqq] {$H^1$};
                \vertex  (h3) [below left = of hqq] {$H^{3}$};
                \vertex (q1) [below right =  of hqq] {$t^5_R$};
                \vertex (q2) [above right =  of hqq] {$Q^4_L$};
                \diagram* {
                    (h)  -- [scalar] (hqq),
                    (q1) -- [anti fermion, line width=\lwR] (hqq) -- [anti fermion, line width=\lwL] (q2),
                    (h2) -- [scalar] (hqq) --[scalar] (h3),
                };
            \end{feynman}
        \end{tikzpicture} &= -i \; \coeff{t H}{*} \left[ 
        \id{i_1 i_2} \varepsilon_{i_3 i_4}+\id{i_3 i_2} \varepsilon_{i_1 i_4}
        \right] \id{m_5 m_4}, 
        \\
        \begin{tikzpicture}[baseline=(hqq)]
            \begin{feynman}[small]
                \vertex  (h)  {$H^{2,\dagger}$};
                \vertex  (hqq) [dot, scale=\sizedot,right= of h] {};
                \vertex  (h2) [above left = of hqq] {$H^1$};
                \vertex  (h3) [below left = of hqq] {$H^{3}$};
                \vertex (q1) [above right =  of hqq] {$b^4_R$};
                \vertex (q2) [below right =  of hqq] {$Q^5_L$};
                \diagram* {
                    (h)  -- [scalar] (hqq),
                    (q1) -- [fermion, line width=\lwR] (hqq) -- [fermion, line width=\lwL] (q2),
                    (h2) -- [scalar] (hqq) --[scalar] (h3),
                };
            \end{feynman}
        \end{tikzpicture} &= +  i\; \coeff{b H }{}\left[ \id{i_1 i_2} \id{i_5 i_3} + \id{i_3 i_2} \id{i_5 i_1} \right] \id{m_5 m_4}  , \\
        \begin{tikzpicture}[baseline=(hqq)]
            \begin{feynman}[small]
                \vertex  (h)  {$H^{2,\dagger}$};
                \vertex  (hqq) [dot, scale=\sizedot,right= of h] {};
                \vertex  (h2) [above left = of hqq] {$H^1$};
                \vertex  (h3) [below left = of hqq] {$H^{3, \dagger}$};
                \vertex (q1) [below right =  of hqq] {$b^5_R$};
                \vertex (q2) [above right =  of hqq] {$Q^4_L$};
                \diagram* {
                    (h)  -- [scalar] (hqq),
                    (q1) -- [anti fermion, line width=\lwR] (hqq) -- [anti fermion, line width=\lwL] (q2),
                    (h2) -- [scalar] (hqq) --[scalar] (h3),
                };
            \end{feynman}
        \end{tikzpicture} &= + i\; \coeff{b H }{*}\left[ \id{i_1 i_2} \id{i_4 i_3} + \id{i_3 i_2} \id{i_4 i_1} \right] \id{m_5 m_4}.
\end{align}
\end{subequations}
To conclude, the Higgs-gluon contact interaction is:
\begin{equation}
\begin{tikzpicture}[baseline=(hhgg)]
            \begin{feynman}[small]
                \vertex  (h1)  {$H^{3,\dagger}$};
                \vertex  (hhgg) [square dot,scale=\sizesqdot,below left= of h1] {};
                \vertex  (h2) [below right = of hhgg] {$H^4$}; 
                \vertex (g1) [above left =  of hhgg] {$g^1$};
                \vertex (g2) [below left =  of hhgg] {$g^2$};
                \diagram* {
                (g1) -- [gluon] (hhgg) -- [gluon] (g2),
                (h1) -- [scalar] (hhgg) -- [scalar] (h2),
                };
            \end{feynman}
        \end{tikzpicture} =  + 4 i \; \coeff{H G}{}  \id{i_3 i_4} \id{A_1 A_2} \left(p_1^{\mu_2}p_2^{\mu_1} - g^{\mu_1 \mu_2} p_1 \cdot p_2 \right).
\end{equation}

\section{Master integrals \label{sec:MI}}
In this Appendix, we describe the computation of the MIs needed for the evaluation of the two loop form factor $\mathcal{A}^{(2)}$. 
For that purpose, we consider two $2$-loop $7$-denominator families of Feynman integrals in $d = 4 - 2 \epsilon$ dimensions of the type 
\begin{equation}
\tau_{a_1,\ldots,a_7} \equiv 
    \int \, \prod _{i=1}^{2} \widetilde{d^{d}l_{i}} \frac{1}{D_{1}^{a_{1}} \cdots D_{7}^{a_{7}}} \ ,
\label{eq:integralfamily}
\end{equation}
for the evaluation of $\mathcal{A}^{(2)}$.
The integration measure $\widetilde{d^{d}l_{i}}$ is defined as
\begin{equation}
  	\label{measure2}
	\widetilde{d^{d}l_{i}} \equiv 
    \frac{d^{d}l_{i}}{i \pi^{d/2}} \, \left(\frac{m^2}{\mu^2}\right)^{\epsilon}\frac{1}{\Gamma(1+\epsilon)} \ .
\end{equation}
where $\mu$ is the 't-Hooft scale of dimensional regularization. We define the two integral families as $T_1$ and $T_2$ corresponding to the representative diagrams as shown in Fig.~\ref{fig:gghh1}, respectively. For the first integral family $T_1$, we choose the following set of propagators
\begin{eqnarray}
    &&D_1 = (l_1)^2, \, \, \, \, D_2 = (l_1 + k_1)^2, \, \, \, \, D_3 = (l_1 - l_2)^2 - m^2, \, \, \, \, D_4 = (l_2 + k_1)^2, \\ \newline
    &&D_5 = (l_2 + k_1 + k_2)^2, \, \, \, \, D_6 = (l_1 \cdot l_2), \, \, \, \, D_7 = (l_1 \cdot k_2) \, ,
\end{eqnarray}
and for the integral family of $T_2$, we choose 
\begin{eqnarray}
    &&D_1 = (l_1)^2, \, \, \, \, D_2 = (l_1 + k_1)^2, \, \, \, \, D_3 = (l_1 + k_1 + k_2)^2, \, \, \, \, D_4 = (l_2 )^2, \\ \newline
    &&D_5 = (l_1 - l_2)^2 - m^2, \, \, \, \, D_6 = (l_2 \cdot k_1), \, \, \, \, D_7 = (l_2 \cdot k_2).
\end{eqnarray}
The $D_6$ and $D_7$ in both the families are considered as irreducible scalar products. Here $l_i$ is the loop momentum and $k_i$ is the external momentum with the kinematics rule $k_1^2 = 0,  k_2^2 = 0$ and $(k_1 + k_2)^2 = s$.

We apply integration-by-parts (IBP) identities on both integral families to generate the differential equation of the master integrals for each family in the dimensionless variable $z = -\frac{s}{m^2}$. Initially, we choose a set of MIs ($\mathbf{F}$) that admits the following system of differential equations:
\begin{equation}
    \frac{\partial \mathbf{F}}{\partial z} = \mathbb{A}(\epsilon, z) \, \mathbf{F},
\end{equation}
where  $\mathbb{A}(\epsilon, z)$ is linear in the dimensional regularization parameter $\epsilon = \frac{4-d}{2}$ and $d$ is the number of space-time dimensions. Now, following the Magnus matrix method~\cite{Argeri:2014qva, DiVita:2014pza}, we find a set of MIs $\mathbf{I}$, which satisfies a canonical systems of differential equations~\cite{Henn:2013pwa}, where the dependence on $\epsilon$ is factorized from the kinematics,
\begin{equation}
    d \, \mathbf{I} = \epsilon \, d \mathbb{A}(z) \, \mathbf{I}.
\end{equation}
The differential equation matrix takes the form
\begin{equation}
d \mathbb{A} (z)= \sum_{i} \mathbb{M}_i \, d \log \left( \eta_i (z) \right),
\label{eq:dA_matrix}
\end{equation}
with the $\mathbb{M}_i$ being the constant matrices with rational entries. The arguments $\eta_i$ form the alphabet of the differential equation, consisting of letters. The MIs are computed in the kinematic region where all the letters are real and positive. This allows us to write the solution of the differential equation as a Taylor series in $\epsilon$:
\begin{equation}
    {\mathbf I} (\epsilon, z) = \sum {\mathbf{I}}^{(j)} (z) \, \,  \epsilon^j.
\end{equation}
The $j-$th order coefficient can be expressed as
\begin{equation}
    \mathbf{I}^{(j)} = \sum_{i=0}^{j} \int_{\gamma} \underbrace{d \mathbb{A}(z) \dots d \mathbb{A}(z)}_{\text{$i$ times}}\,  \mathbf{I}^{(j-i)}(z_0)\ .
\label{eq:solution_DEQ_implicit}
\end{equation}
where $\gamma$ denotes a regular path in the $z$-plane, and $\mathbf{I}^{(j-i)}(z_0)$ are boundary constants. Following Eq.~(\ref{eq:dA_matrix}), $\mathbf{I}^{(j)}$ can be expressed in 
terms of Harmonic polylogarithms (HPLs)~\cite{REMIDDI_2000,Gehrmann_2001} up to weight 3.
%
%
Now, we explicitly show the results for the integral families $T_1$ and $T_2$.
\paragraph{Family $T_1$ :}
We choose the following set of MIs, which satisfy an $\epsilon$ linear differential equation:
\begin{equation}
    \text{F}_1 = \epsilon^2 (\epsilon - 1) \tau_{0, 2, 1, 0, 1, 0, 0} \, ,  \hspace{0.3cm} 
    \text{F}_2 = \epsilon^2 \tau_{1, 0, 2, 0, 2, 0, 0} \, ,
    \hspace{0.3cm} 
    \text{F}_3 = \epsilon^2 \tau_{2, 0, 1, 0, 2, 0, 0} \, , 
    \hspace{0.3cm} 
    \text{F}_4 = \epsilon^3 \tau_{1, 1, 2, 1, 1, 0, 0} \, . 
\end{equation}
The MIs ($\tau$) can be pictorially represented as follows. We use a thick line to denote the massive propagator and a thin line to denote the massless propagators. The dot denotes a double insertion of the propagator. 
\begin{equation}
\begin{aligned}
\tau_{0,2,1,0,1,0,0} &= \begin{tikzpicture}[baseline=(k1)]
            \begin{feynman}[small]
                \vertex  (k1)  {};
                \vertex  (v1) [right= 20 pt of k1, scale=0.01]  {};
                \vertex  (v2) [above right= of v1, dot, scale=1.5]  {};
                \vertex  (v3) [below right= of v1, scale=.01]  {};
                \vertex  (v4) [below right= of v2, scale=.01]  {};
                \vertex (k2) [right = 20 pt  of v4] {};
                \diagram* {
                (k1) -- [] (v1) -- [quarter left] (v2) --  [quarter left] (v4) -- (k2),
                (v4) -- [quarter left] (v3) -- [quarter left] (v1), 
                (v1) -- [line width = 1.5] (v4),
                };
            \end{feynman}
        \end{tikzpicture}, \quad 
\tau_{1,0,2,0,2,0,0} = \begin{tikzpicture}[baseline=(v1)]
            \begin{feynman}[small]
                \vertex  (k11)  {};
                \vertex  (v1) [below right= of k11, scale=0.01]  {};
                \vertex  (k12) [below left= of v1]  {};
                \vertex  (v2) [above right= of v1, dot, scale=1.5]  {};
                \vertex  (v3) [below right= of v1, scale=.01]  {};
                \vertex (v5) [right = 20 pt of v1, dot, scale=1.5] {};
                \vertex  (v4) [below right= of v2,dot,scale=.01]  {};
                \vertex (k21) [above right =  of v4] {};
                \vertex (k22) [below right =  of v4] {};
                \diagram* {
                (k11) -- [] (v1) -- [quarter left] (v2) --  [quarter left] (v4) -- (k21),
                (v4) -- (k22),
                (v1) -- (k12),
                (v4) -- [quarter left] (v3) -- [quarter left] (v1), 
                (v1) -- [line width = 1.5] (v4),
                };
            \end{feynman}
        \end{tikzpicture}, \\ 
 \tau_{2,0,1,0,2,0,0} &=  \begin{tikzpicture}[baseline=(v1)]
            \begin{feynman}[small]
                \vertex  (k11)  {};
                \vertex  (v1) [below right= of k11, scale=0.01]  {};
                \vertex  (k12) [below left= of v1]  {};
                \vertex  (v2) [above right= of v1, dot, scale=1.5]  {};
                \vertex  (v3) [below right= of v1, dot,scale=1.5]  {};
                \vertex (v5) [right = 20pt of v1, dot, scale=0.01] {};
                \vertex  (v4) [below right= of v2,dot,scale=.01]  {};
                \vertex (k21) [above right =  of v4] {};
                \vertex (k22) [below right =  of v4] {};
                \diagram* {
                (k11) -- [] (v1) -- [quarter left] (v2) --  [quarter left] (v4) -- (k21),
                (v4) -- (k22),
                (v1) -- (k12),
                (v4) -- [quarter left] (v3) -- [quarter left] (v1), 
                (v1) -- [line width = 1.5] (v4),
                };
            \end{feynman}
        \end{tikzpicture}, \quad 
\tau_{1,1,2,1,1,0,0} = \begin{tikzpicture}[baseline=(v2)]
            \begin{feynman}[small]
                \vertex  (k11) [] {};
                \vertex  (v1) [right = of k11, scale=0.01]  {};
                \vertex  (v2) [below right = of v1, dot, scale=0.01]  {};
                \vertex (v3) [below left= of v2, scale=.01]  {};
                \vertex (v4) [below = 20 pt of v1, dot, scale = 0.01] {};
                \vertex (v5) [right = 10 pt of v4, dot, scale = 1.5] {};
                \vertex (k12) [left = of v3]  {};
                \vertex (k21) [above right = of v2] {};
                \vertex (k22) [below right = of v2] {};
                \diagram* {
                (k21) -- (v2) -- (k22),
                (k11) -- (v1) -- (v2) -- (v3) -- (k12),
                (v1) -- (v3),
                (v4) -- [line width = 1.5] (v2)
                };
            \end{feynman}
        \end{tikzpicture}.
\end{aligned}
\end{equation}
Subsequently, we use the Magnus exponential method to find a new basis of MIs $\text{I}_i$ that satisfy the canonical differential equation in $z$:
\begin{equation}
    \text{I}_1 = \text{F}_1,  \hspace{0.5cm} \text{I}_2 = - s \, \text{F}_2, \hspace{0.8cm}  
    \text{I}_3 = 2 \, m^2 \text{F}_2 + (m^2 - s) \text{F}_3, \hspace{0.8cm} \text{I}_4 = - m^2 \, s \, \text{F}_4, 
\end{equation}
The canonical differential equation matrix takes the form as shown in Eq.~(\ref{eq:dA_matrix}), where we get two matrices $\mathbb{M}_1$, and $\mathbb{M}_2$ having rational number as entries and the letters are
\begin{equation}
    \eta_1 = z, \hspace{1cm} \eta_2 = (1+ z)
\end{equation}
This system of the differential equations are easily integrated to HPLs using \texttt{PolyLogTools}~\cite{Duhr:2019tlz} and numerically evaluated by \texttt{GiNaC}~\cite{Vollinga:2004sn}.
We determine the boundary constants by using the numerical evaluation of these integrals from the package {\texttt{AMFlow}}~\cite{Liu:2022chg}, based on the auxiliary mass flow method, at the point $s = -59, m^2 = 9$ with 100 digits precision. Later, these boundary constants are analytically reconstructed using the {\texttt{PSLQ}} algorithm~\cite{Ferguson1999}.
The final expression of the MIs are,
\begin{align}
\text{I}_1 &= \frac{1}{2} + \epsilon^2 \, \zeta(2) - \epsilon^3 \, \zeta(3) + \mathcal{O}\left(\epsilon^4\right),
\\
\begin{split}
\text{I}_2 &= \epsilon \, H(-1,z) 
+ \epsilon^2 \Bigg[ H(0,-1,z) - 4 \, H(-1,-1,z) \Bigg]  \\
& + \epsilon^3 \Bigg[ 2 \, \zeta(2) \, H(-1,z) + 16 \, H(-1,-1,-1,z) 
- 4 \, H(0,-1,-1,z)   \\
& \quad \qquad - 6 \, H(-1,0,-1,z)  + H(0,0,-1,z) \Bigg] + \mathcal{O}\left(\epsilon^4\right),
\end{split} \\
\begin{split}
\text{I}_3 &= -1 + 2 \, \epsilon \, H(-1,z) 
+ \epsilon^2 \Bigg[ -8 \, H(-1,-1,z) + 4 \, H(0,-1,z) - 2 \, \zeta(2) \Bigg] \\
& + \epsilon^3 \Bigg[ 4 \, \zeta(2) \, H(-1,z) + 32 \, H(-1,-1,-1,z) - 12 \, H(-1,0,-1,z) 
  \\
& \qquad \quad - 16 \, H(0,-1,-1,z)+ 4 \, H(0,0,-1,z) + 2 \, \zeta(3) \Bigg] + \mathcal{O}\left(\epsilon^4\right),
\end{split} \\ 
\begin{split}
\text{I}_4 &= \epsilon \, H(-1,z) 
+ \epsilon^2 \Bigg[ 4 \, H(0,-1,z) - 4 \, H(-1,-1,z) \Bigg]\\
& + \epsilon^3 \Bigg[ 2 \, \zeta(2) \, H(-1,z) + 16 \, H(-1,-1,-1,z) - 6 \, H(-1,0,-1,z) 
 \\
& \quad \qquad - 16 \, H(0,-1,-1,z)  + 6 \, H(0,0,-1,z) \Bigg] + \mathcal{O}\left(\epsilon^4\right).
\end{split}
\end{align}

\paragraph{Family $T_2$ :}
For the integral family $T_2$, we choose the following set of MIs, which satisfy an $\epsilon$ linear differential equation:
\begin{equation}
    \text{F}_1 = \epsilon^2 \tau_{0, 0, 2, 1, 2, 0, 0} \, ,  
    \hspace{0.3cm} 
    \text{F}_2 = \epsilon^2 \tau_{0, 0, 2, 2, 1, 0, 0} \, , 
    \hspace{0.3cm}  
    \text{F}_3 = \epsilon^2 (\epsilon -1) \tau_{0, 2, 0, 1, 1, 0, 0} \, , 
    \hspace{0.3cm} 
    \text{F}_4 = \epsilon^2 \tau_{2, 0, 1, 0, 2, 0, 0}.
\end{equation}
The MIs can be pictorially represented as follows (following the same notation as before):
\begin{equation}
\begin{aligned}
\tau_{0,0,2,1,2,0,0} &= \begin{tikzpicture}[baseline=(v1)]
            \begin{feynman}[small]
                \vertex  (k11)  {};
                \vertex  (v1) [below right= of k11, scale=0.01]  {};
                \vertex  (k12) [below left= of v1]  {};
                \vertex  (v2) [above right= of v1, dot, scale=1.5]  {};
                \vertex  (v3) [below right= of v1, scale=.01]  {};
                \vertex (v5) [right = 20 pt of v1, dot, scale=1.5] {};
                \vertex  (v4) [below right= of v2,dot,scale=.01]  {};
                \vertex (k21) [above right =  of v4] {};
                \vertex (k22) [below right =  of v4] {};
                \diagram* {
                (k11) -- [] (v1) -- [quarter left] (v2) --  [quarter left] (v4) -- (k21),
                (v4) -- (k22),
                (v1) -- (k12),
                (v4) -- [quarter left] (v3) -- [quarter left] (v1), 
                (v1) -- [line width = 1.5] (v4),
                };
            \end{feynman}
        \end{tikzpicture}, \quad 
\tau_{0,0,2,2,1,0,0} = \begin{tikzpicture}[baseline=(v1)]
            \begin{feynman}[small]
                \vertex  (k11)  {};
                \vertex  (v1) [below right= of k11, scale=0.01]  {};
                \vertex  (k12) [below left= of v1]  {};
                \vertex  (v2) [above right= of v1, dot, scale=1.5]  {};
                \vertex  (v3) [below right= of v1, dot, scale = 1.5]  {};
                \vertex (v5) [right = 20 pt of v1, dot, scale=.01] {};
                \vertex  (v4) [below right= of v2,dot,scale=.01]  {};
                \vertex (k21) [above right =  of v4] {};
                \vertex (k22) [below right =  of v4] {};
                \diagram* {
                (k11) -- [] (v1) -- [quarter left] (v2) --  [quarter left] (v4) -- (k21),
                (v4) -- (k22),
                (v1) -- (k12),
                (v4) -- [quarter left] (v3) -- [quarter left] (v1), 
                (v1) -- [line width = 1.5] (v4),
                };
            \end{feynman}
        \end{tikzpicture}, \\ 
 \tau_{0,2,0,1,1,0,0} &= \begin{tikzpicture}[baseline=(k1)]
            \begin{feynman}[small]
                \vertex  (k1)  {};
                \vertex  (v1) [right= 20 pt of k1, scale=0.01]  {};
                \vertex  (v2) [above right= of v1, dot, scale=1.5]  {};
                \vertex  (v3) [below right= of v1, scale=.01]  {};
                \vertex  (v4) [below right= of v2, scale=.01]  {};
                \vertex (k2) [right = 20 pt  of v4] {};
                \diagram* {
                (k1) -- [] (v1) -- [quarter left] (v2) --  [quarter left] (v4) -- (k2),
                (v4) -- [quarter left] (v3) -- [quarter left] (v1), 
                (v1) -- [line width = 1.5] (v4),
                };
            \end{feynman}
        \end{tikzpicture}, \quad 
\tau_{2,0,1,0,2,0,0} = \begin{tikzpicture}[baseline=(v1)]
            \begin{feynman}[small]
                \vertex  (k11)  {};
                \vertex  (v1) [below right= of k11, scale=0.01]  {};
                \vertex  (k12) [below left= of v1]  {};
                \vertex  (v2) [above right= of v1, dot, scale=.01]  {};
                \vertex  (v3) [right= 20 pt of v1, dot,  scale=1.5]  {};
                \vertex  (v4) [below right= of v2,scale=.01, dot]  {};
                \vertex (k21) [ above right =  of v4] {};
                \vertex (k22) [ right = 20 pt of v4] {};
                \vertex (v5) [below = 12 pt of v4, dot, scale=1.5] {};
                \diagram* {
                (k11) -- [] (v1) -- [quarter left] (v2) --  [quarter left] (v4) -- (k21),
                (v4) -- (k22),
                (v1) -- (k12),
                (v4) --  (v3) -- (v1), 
                (v4) -- [half right, line width = 1.5] (v5) -- [half right, line width = 1.5] (v4),
                };
            \end{feynman}
        \end{tikzpicture} 
        .
\end{aligned}
\end{equation}
Subsequently, we use the Magnus exponential method to find a new basis of MIs $I_i$ that satisfy the canonical differential equation in $z$:
\begin{equation}
   \text{I}_1 = - s \, \text{F}_1,  \hspace{1cm} \text{I}_2 = 2\, m^2 \text{F}_1 + (m^2 -s) \text{F}_2 \hspace{1cm}  \text{I}_3 = \text{F}_3, \hspace{1cm} \text{I}_4 = - s \, \text{F}_4, 
\end{equation}
The canonical differential equation matrix takes the form as shown in Eq.~(\ref{eq:dA_matrix}), where we get two matrices $\mathbb{M}_1$, and $\mathbb{M}_2$ having rational number as entries and the letters are, as for the integral family $T_1$, $\eta_1 = z$, and  $\eta_2 = (1+ z)$.
Using the same procedure for integrating the differential equation matrix and finding the boundary constant as outlined for the integral family $T_1$, the final expressions read
\begin{align}
\begin{split}
\text{I}_1 &= \epsilon \, H(-1,z) 
+ \epsilon^2 \Bigg[ H(0,-1,z) - 4 \, H(-1,-1,z) \Bigg]   \\
& + \epsilon^3 \Bigg[ 2 \, \zeta(2) \, H(-1,z) + 16 \, H(-1,-1,-1,z) - 6 \, H(-1,0,-1,z) 
  \\
&\quad \qquad - 4 \, H(0,-1,-1,z) + H(0,0,-1,z) \Bigg] + \mathcal{O}\left(\epsilon^4\right) \, ,
\end{split}
\\
\begin{split}
\text{I}_2 &= -1 + 2 \, \epsilon \, H(-1,z) 
+ \epsilon^2 \Bigg[ -8 \, H(-1,-1,z) + 4 \, H(0,-1,z) - 2 \, \zeta(2) \Bigg]  \\
& + \epsilon^3 \Bigg[ 4 \, \zeta(2) \, H(-1,z) + 32 \, H(-1,-1,-1,z) - 12 \, H(-1,0,-1,z)  \\
& \qquad \quad - 16 \, H(0,-1,-1,z) + 4 \, H(0,0,-1,z) + 2 \, \zeta(3) \Bigg] + \mathcal{O}\left(\epsilon^4\right) \, ,    
\end{split}
\\
\begin{split}
\text{I}_3 &= \frac{1}{2} + \epsilon^2 \, \zeta(2) 
- \epsilon^3 \, \zeta(3) + \mathcal{O}\left(\epsilon^4\right) \, ,    
\end{split}
\\
\begin{split}
\text{I}_4 &= 1 - \epsilon \, H(0,z) + \epsilon^2 \Bigg[ H(0,0,z) - \zeta(2) \Bigg]  \\
& + \epsilon^3 \Bigg[ \zeta(2) \, H(0,z) - H(0,0,0,z) - 2 \, \zeta(3) \Bigg]  +\mathcal{O}\left(\epsilon^4\right) \, .    
\end{split}
\end{align}

{To the best of our knowledge the top-level topology integrals discussed here are provided for the first time in the literature. We finally note for our purpose the determination in the Euclidean region is sufficient. Furthermore, as previously mentioned, it is crucial to address the source of the divergences carefully. In our calculations, additional massless integrals emerge, but these can be expressed as products of one-loop integrals available in standard literature. }

\clearpage
\newpage

\bibliographystyle{utphys.bst}
\bibliography{bibliography}

\end{document}